\documentclass[10pt, two column, twoside]{IEEEtran}
\usepackage{color}
\usepackage[colorlinks, linkcolor=color1, anchorcolor=blue, citecolor=color1]{hyperref}
\usepackage[OT1]{fontenc} 

\usepackage{hyperref}
\usepackage{enumitem}

\usepackage{amssymb}
\usepackage{amsmath}
\usepackage[lined,boxed,commentsnumbered, ruled]{algorithm2e}
\usepackage{mathrsfs}
\usepackage{algorithmic}
\usepackage{bm}
\usepackage{array}
\newcolumntype{C}[1]{>{\centering\arraybackslash}p{#1}}
\newcolumntype{M}[1]{>{\centering\arraybackslash}m{#1}}
\usepackage{subfigure}
\usepackage{graphicx,booktabs,multirow}

\definecolor{colorhkust}{RGB}{20,43,140}
\definecolor{colortsinghua}{RGB}{116,52,129}
\definecolor{color1}{RGB}{128,0,0}

\newcommand{\rank}{\mathrm{rank}}

\date{}

\begin{document}

\title{{Communication-Efficient Edge AI: Algorithms and Systems}}
\author{Yuanming~Shi,~\IEEEmembership{Member,~IEEE,}~Kai~Yang,~\IEEEmembership{Student Member,~IEEE,}~Tao~Jiang,~\IEEEmembership{Student Member,~IEEE,} Jun~Zhang,~\IEEEmembership{Senior Member,~IEEE,}~and~Khaled~B.~Letaief,~\IEEEmembership{Fellow,~IEEE}

\thanks{Y. Shi and T. Jiang are with the School of Information Science and Technology, ShanghaiTech University, Shanghai 201210, China (e-mail: \{shiym,jiangtao1\}@shanghaitech.edu.cn).}
\thanks{K. Yang is with the School of Information Science and Technology, ShanghaiTech University, Shanghai 201210, China, also with the Shanghai Institute of Microsystem and Information Technology, Chinese Academy of Sciences, Shanghai 200050, China, and also with the University of Chinese Academy of Sciences, Beijing 100049, China (e-mail: yangkai@shanghaitech.edu.cn).}
\thanks{J. Zhang is with the Department of Electronic and Information Engineering, The Hong Kong Polytechnic University, Hong Kong (E-mail: jun-eie.zhang@polyu.edu.hk).}
\thanks{K. B. Letaief is with the Department of Electronic and
Computer Engineering, Hong Kong University of Science and Technology,
Hong Kong (E-mail: eekhaled@ust.hk). He is also with Peng Cheng Laboratory, Shenzhen, China.}
}                       
\maketitle
\IEEEpeerreviewmaketitle


\maketitle

\begin{abstract}
Artificial intelligence (AI) has achieved remarkable breakthroughs in a wide range of fields, ranging from speech processing, image classification to drug discovery. This is driven by the explosive growth of data, advances in machine learning (especially deep learning), and easy access to vastly powerful computing resources. Particularly, the wide scale deployment of edge devices (e.g., IoT devices) generates an unprecedented scale of data, which provides the opportunity to derive accurate models and develop various intelligent applications at the network edge. 
However, such enormous data cannot all be sent from end devices to the cloud for processing, due to the varying channel quality, traffic congestion and/or privacy concerns. 
By pushing inference and training processes of AI models to edge nodes, edge AI has emerged as a promising alternative. AI at the edge requires close cooperation among \textit{edge devices}, such as smart phones and smart vehicles, and \textit{edge servers} at the wireless access points and base stations, which however result in heavy communication overheads. In this paper, we present a comprehensive survey of the recent developments in various techniques for overcoming these communication challenges. Specifically, we first identify key communication challenges in edge AI systems. We then introduce communication-efficient techniques, from both algorithmic and system perspectives for training and inference tasks at the network edge. Potential future research directions are also highlighted.
\end{abstract}


\section{Introduction}
With the explosive growth in data and the rapid advancements of algorithms (e.g., deep learning), as well as the step-change improvement of computing resources, artificial intelligence (AI)  has achieved breakthroughs in a wide range of applications, including  speech processing \cite{graves2013speech}, image classification \cite{krizhevsky2012imagenet} and reinforcement learning \cite{arulkumaran2017deep}, etc. AI is expected to affect significant segments of many vertical industries and our daily life, such as intelligent vehicles \cite{zhang2019mobile} and tactile robots \cite{haddadin2018tactile}. In addition, it is anticipated that AI could add around 16 percent or about $\$13$ trillion to the global gross domestic product (GDP) by 2030, compared with that of 2018 \cite{Bughin2018}. 

The explosive data growth generated by the massive number of end devices, e.g., smart phones, tablets and Internet-of-Things (IoT) sensors,  provides opportunities and challenges for providing intelligent services. It is predicted that there will be nearly 85 Zettabytes of usable data generated by all people, machines and things by 2021, which shall exceed the cloud data center traffic (21 Zettabytes) by a factor of \(4\) \cite{Cisco2018}.  Moreover, delay-sensitive intelligent applications, such as autonomous driving, cyber-physical control systems, and robotics, require fast processing of the incoming data. Such extremely high network bandwidth and low latency requirements would place  unprecedented pressures on traditional cloud-based AI, where massive sensors/embedded devices transfer collected data to the cloud \cite{sze2017efficient}, often under varying network qualities (e.g., bandwidth and latency).  In addition, privacy is a major concern for cloud-based solutions.
To address these problems, one promising solution, edge AI \cite{zhou2019edge,park2019wireless}, comes to the rescue. 

Futuristic wireless systems \cite{letaief2019roadmap} mainly consist of ultra-dense edge nodes, including \textit{edge servers} at the base stations and wireless access points, and \textit{edge devices} such as smart phones, smart vehicles, and drones. Edge AI pushes inference and training processes of AI models to the network edge in close proximity to data sources. As such, the amount of data transferred to the cloud will be significantly reduced, thus alleviating the network traffic load, latency and privacy concerns. Although training an AI model (e.g., deep neural networks) generally requires intensive computing resources, the rapid development of mobile edge computing can provide cloud-computing capabilities at the edge of the mobile network \cite{hu2015mobile,Khaled_MEC17}, and make the application of AI to edges much more efficient. In addition, computational capabilities of edge servers and edge devices continue to improve. Notable examples include the deployment of neural processing unit (NPU) in Kirin 970 smart phone chips and the Apple's bionic chip A12, which substantially accelerate AI computations on edge devices. In a nutshell, the advances of mobile edge computing platforms and the improvement of computing power of the edge nodes make edge AI a feasible solution.
\begin{figure*}[ht]
      \centering
      \includegraphics[width=0.8\linewidth]{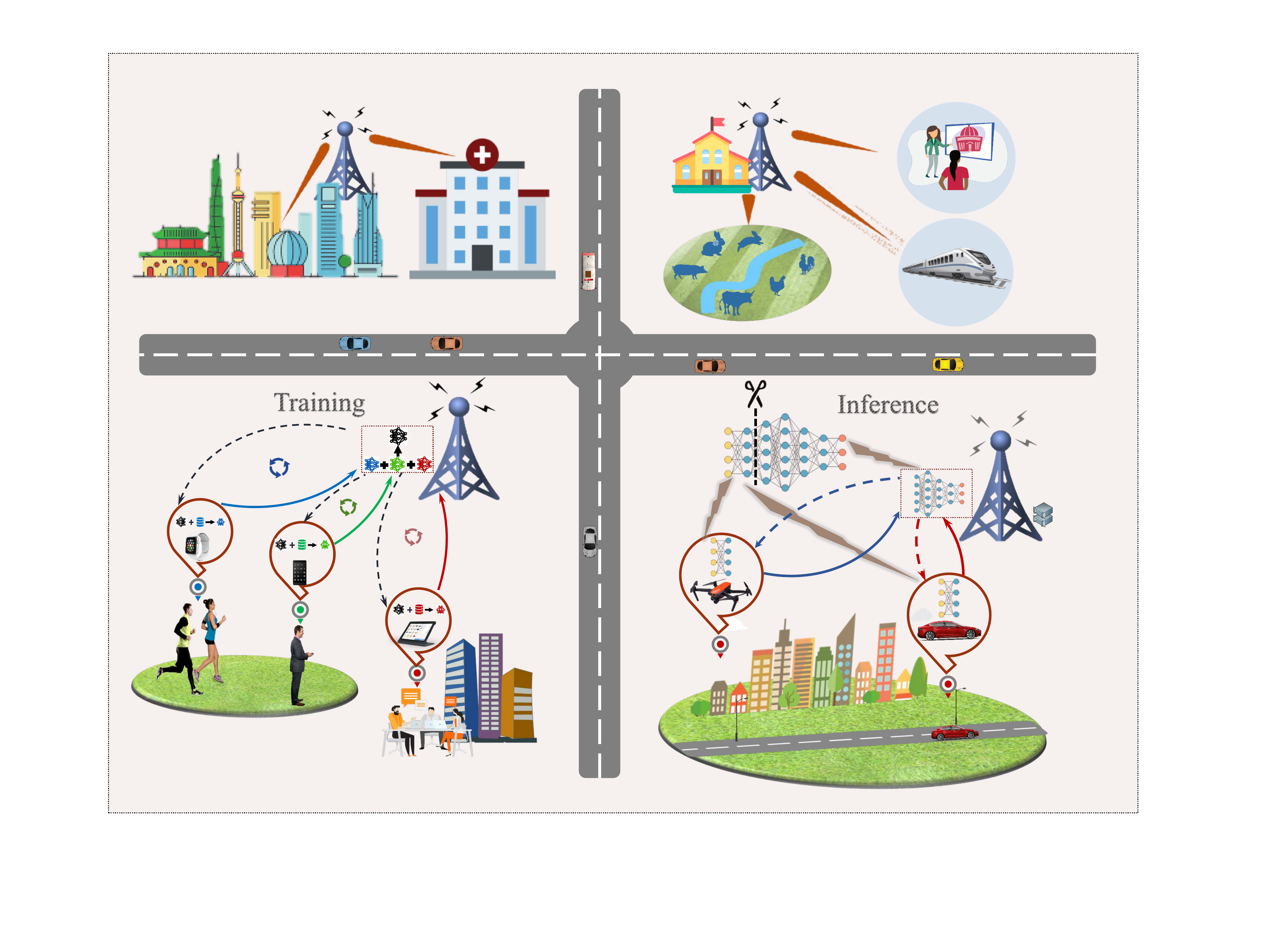}
      \caption{Illustration of edge AI including edge training and edge inference.}
      \label{fig:system}
\end{figure*} 

Nevertheless, pushing AI towards the edge is a non-trivial task. The most straightforward way of realizing edge AI without any communication load, i.e., deploying the full AI models on edge devices, is often infeasible when the size of the AI model (e.g., DNNs) is too large or the computational requirement is too high, given the limited hardware resources of edge devices.  
A promising solution is to incorporate cooperation among edge nodes to accomplish edge AI tasks that require intensive computation and large storage sizes. This can be achieved by exploiting different data storage and processing capabilities for a wider range of intelligent services with distinct latency and bandwidth requirements \cite{zhu2018towards}, as shown in Fig. \ref{fig:system}. For example, based on federated learning \cite{konevcny2015federated}, we can use multiple devices to train an AI model collaboratively. Specifically, each device only needs to compute a local model according to its own data samples, before sending the computation results to a fusion center, where the global AI model is aggregated and updated. The new AI model will be transmitted back to each device for training at the next epoch. Such solutions exploit on-device computing power in a collaborative way, which, however, requires significant communication overheads during the model updating process. In addition, some computation-intensive AI inference tasks can only be accomplished by task-splitting among edge devices and edge servers \cite{kang2017neurosurgeon}, which also incurs heavy communication cost. Therefore, the enormous communication overhead presents a major bottleneck for edge AI.

To unleash its full potential, the upcoming edge AI \cite{zhou2019edge,li2019federated} shall rely on advances in various aspects, including the smart design of distributed learning algorithms and system architectures, supported by efficient communication protocols. In this article, we survey the communication challenges in designing and deploying AI models and algorithms in edge AI systems. Specifically, we provide a thorough survey on  communication-efficient distributed learning algorithms for training AI models on edges. In addition, we provide an overview of edge AI system architectures for communication-efficient edge training and edge inference. In the next section, we start with the motivations and identify major communication challenges in edge AI. A paper outline will also be provided.

\section{Motivations and Challenges}
In this section, we present the motivations and identify key communication challenges of edge AI. Interplays among computation mechanisms, learning algorithms, as well as system architectures, are revealed.

\subsection{Motivations}
During the past decades, the thriving mobile internet has enabled various mobile applications such as mobile pay, mobile gaming, etc. These applications in turn led to an  upsurge of mobile devices and mobile data, which prompts the prosperity of AI for greatly facilitating daily life. As the key design target, the 5G roll-out has focused on several key services for \textit{connected things}: enhanced mobile broadband (eMBB),  ultra reliable low latency communications (URLLC), and massive machine type communications (mMTC). In contrast, futuristic 6G networks \cite{letaief2019roadmap} will undergo a paradigm shift from \textit{connected things} to \textit{connected intelligence}. The network infrastructure of 6G is envisioned to fully exploit the potential of massive distributed devices and the data generated at network edges for supporting intelligent applications \cite{zhou2019edge}.

In recent years, a new trend is to move computation tasks from the cloud center towards network edges due to the increasing computing power of edge nodes \cite{Khaled_MEC17}. In the upcoming 5G wireless systems, there are a growing number of edge nodes, varying from base stations (BSs) to various edge devices such as mobile phones, tablets, and IoT devices. The computational capabilities of edge mobile devices have seen substantial improvements thanks to the rapid development of mobile chipsets. For example, mobile phones nowadays have comparable computing power as computing servers a decade ago. In addition, edge servers have the potential to provide low-latency AI services for mobile users which are infeasible to be directly implemented on devices. Since edge servers have relatively less powerful computation resources than cloud centers, it is necessary to employ joint design principles across edge servers and edge devices to further reduce execution latency and enhance privacy \cite{Khaled_MEC17}. The advances in edge computing thus provides opportunities for pushing AI frontiers from the cloud center to network edges, stimulating a new research area known as \textit{edge AI}, including both AI model training and inference procedures.

Training at network edges is challenging and requires coordinating massive edge nodes to collaboratively build a machine learning model \cite{mcmahan2017communication}. Each edge node usually has access to only a small subset of training data, which is the fundamental difference from traditional cloud-based model training \cite{yang2019federated}. The information exchange across edge nodes in edge training results in high communication cost, especially in the wireless environment with limited bandwidth. This brings a main bottleneck in edge training. It is interesting to note that a number of works have revisited the communication theory for addressing the communication challenges of edge training. The connection between data aggregation from distributed nodes in edge training and the in-network computation problem \cite{giridhar2006toward} in wireless sensor networks has been established in \cite{yang2018federated}, which proposed an over-the-air computation approach for fast model aggregation in each round of training for on-device federated learning. In wireless communication systems, limited feedback \cite{love2008overview} from a receiver to a transmitter is critical to reducing the information bits for realizing channel agile techniques that require channel knowledge at the transmitter. A connection between limited feedback in wireless communication and the quantization method was established in \cite{du2019high} for reducing the data transmission cost in edge training, which borrows ideas from the widely adopted Grassmannian quantization approach for limited feedback.

Edge inference, i.e., performing inference of AI models at network edges, enjoys the benefits of low-latency and enhanced privacy, which are critical for a wide range of AI applications such as drones, smart vehicles, and so on. As such, it has drawn significant attention from both academia and industry. Recently, deep learning models have been actively adopted in a number of applications to provide high-quality services for mobile users. For example, AI technologies have shown promises in healthcare \cite{reddy2019artificial}, such as detection of heart failure \cite{choi2016using} with recurrent neural network (RNN) and decisions about patient treatment \cite{gottesman2019guidelines} with reinforcement learning. However, deep neural network (DNN) models often have a huge number of parameters, which will consume considerable storage and computation resources. A typical example is the classic convolutional neural network (CNNs) architecture named AlexNet \cite{krizhevsky2012imagenet}, which has over 60 million parameters. Therefore, model compression approaches \cite{han2015deep,cheng2018model} have attracted much attention for deploying DNN models at network edges. It should also be noted that the power budget on edge devices is also limited, which stimulates research pursuits on energy-efficient processing of deep neural networks from signal processing perspective \cite{sze2017efficient}. For IoT devices without enough memory to store the entire model, coding techniques shed light on the efficient data shuffling for distributed inference across edge nodes \cite{Ali_arXiv16WDC,Yuanming_WDCTSP18}.

\subsection{Performance Measurements and Unique Challenges of Edge AI}
The typical procedures for providing an AI service include \textit{training} a machine learning model from data, and performing \textit{inference} with the trained model. 
The performance of a machine learning model can be measured by its model accuracy, which can potentially be improved by collecting more training data. However, training a machine learning model from massive data is time consuming. To train a model efficiently, distributed architectures are often adopted, which will introduce additional communication costs for exchanging information across nodes. The computation and communication costs grow extremely high for high-dimensional models such as deep neural networks. In addition, low-latency is also critical for inference in applications such as smart vehicles, smart drones, etc. We thus summarize the key performance measurements of edge AI in terms of \textbf{model accuracy} and \textbf{total latency}.

In the cloud center, cloud computing servers are connected with extremely high bandwidth networks and the training data is available to all nodes. Fundamentally distinct from cloud based AI, edge AI poses more stringent constraints on the algorithms and system architectures. 

\begin{itemize}
  \item \textbf{Limited resources on edge nodes:} Instead of the large amount of powerful GPUs and CPUs integrated servers at the nodes of cloud-based AI, there are often limited computation, storage, and power resources on edge devices, with limited link bandwidth among a large number of edge devices and the edge servers at base stations and wireless access points. For example, the classic AlexNet \cite{krizhevsky2012imagenet}, which is designed for computer vision, has over 60 million parameters. With 512 Volta GPUs interconnected at the rate of 56Gbps, the Alexnet can be trained within record of 2.6 minutes in the data center of SenseTime \cite{sun2019optimizing}. As one of the most powerful GPUs in the world, one Volta GPU has 5,120 cores. However, the Mali-G76 GPU on Huawei Mate 30 Pro, one of the most powerful smart phones, has only 16 cores. The theoretical maximal speed envisioned in 5G is 10Gbps and the average speed is only 50Mbps.

  \item \textbf{Heterogeneous resources across edge nodes:} The variabilities in hardware, network, and power budget of edge nodes imply heterogeneous communication, computation, storage and power capabilities. The edge servers at base stations have much more computation, storage and power resources than mobile devices. For example, Apple Watch Series 5 can only afford up to 10 hours of audio playback\footnote{\url{https://www.apple.com/ca/watch/battery/}}, and users may want to be involved in training tasks only when the devices are charged. To make things worse, edge devices that are connected to a metered cellular network are usually not willing to exchange information with other edge nodes.

  \item \textbf{Privacy and security constraints:} The privacy and security of AI services are increasingly vital especially for emerging high-stake applications in intelligent IoT. Operators expect stricter regulations and laws on preserving data privacy for service providers. For example, the General Data Protection Regulation (GDPR) \cite{euGDPR2016} by the European Union grants users the right for data to be deleted or withdrawn. Federated learning \cite{konevcny2015federated,yang2019federated} becomes a particular relevant research topic for collaboratively building machine learning models while preserving data privacy. Robust algorithms and system designs are also proposed in \cite{Chen_Byzantine17,dong2019secure} for security concern against adversarial attacks during distributed edge training.
\end{itemize}

Enabling efficient edge AI is challenging for coordinating and scheduling edge nodes to efficiently perform a training or inference task under various physical and regulatory constraints. To provide efficient AI services, we shall jointly design new distributed paradigms for computing, communications, and learning. Note that the communication cost for cloud-based AI services may be relatively small compared with computational cost. However, in edge AI systems, the communication cost often becomes a dominating issue due to the stringent constraints. This paper will give a comprehensive survey on edge AI from the perspective of addressing communication challenges from both the algorithm level and system level.

\subsection{Communication Challenges of Edge AI}
Generally, there are multiple communication rounds between edge nodes for an edge AI task. Let $L$ denote the total size of information to be exchanged per round, $r$ denote the communication rate, $N$ denote the number of communication rounds, and $T$ denote the total computation time. Then the total latency in an edge AI system is given by 
\begin{equation}
  \text{Latency} = \underbrace{L/r\cdot N}_{\text{communication}}+\underbrace{\vphantom{L/r}T\cdot N}_{\text{computation}}.
 \end{equation}
For model training, iterative algorithms are often adopted which involve multiple communication rounds. The inference process often requires one round of collaborative computations across edge nodes. Therefore, to alleviate the communication overheads under resource and privacy constraints, it is natural to seek methods for reducing the number of communication rounds for training and the communication overhead per round for training and inference, as well as improving the communication rate. 


From the end-to-end data transmission perspective, the information content of a message is measured in entropy that characterizes the amount of uncertainty. Based on this measure, the limit of lossless source coding is characterized by Shannon's source coding theory \cite{shannon1948mathematical}. It provides a perfect answer to the best we can do if we only focus on ``how to transmit'' instead of ``what to transmit'' from one node to another. That is, the fundamental limit of the end-to-end communication problem has already been solved when the edge AI system and algorithm are fixed. 

However, communication is not isolated in edge AI. From the learning algorithm perspective, ``what to transmit'' determines the required communication overhead per round and the number of communication rounds. This learning level perspective motivates the development of different algorithms to reduce the communication overhead per round and improve the convergence rate. For instance, many gradient based algorithms have been proposed for accelerating the convergence of distributed training \cite{yuan2018variance,lee2017distributed2} . In addition, lossy compression techniques such as quantization and pruning \cite{han2015deep,lin2018deep} have drawn much attention recently to reduce the communication overhead per round.

Edge AI system design has also a great influence on the communication paradigm design across edge nodes. For instance, the target of communication in each round is to compute a certain function value with respect to the intermediate values at edge devices. In particular, the full gradient can be computed at a centralized node by aggregating the locally computed partial gradients at all local nodes. It is therefore better to be studied from the perspective of in-network computation \cite{giridhar2006toward}, instead of treating communication and computation separately. For example, an over-the-air computation approach was developed in \cite{yang2018federated} for fast model aggregation in distributed model training for ferderated learning. In addition, efficient inference at network edges is closely related to computation offloading in edge computing \cite{Khaled_MEC17}, which is being extensively studied in both the communication and mobile computing communities. 

\subsection{Related Works and Our Contributions}
There exist a few survey papers \cite{zhou2019edge,park2019wireless,murshed2019machine,deng2019edge} on edge AI. Particularly, the early works \cite{zhou2019edge,park2019wireless} emphasized the differences between cloud-based AI and edge AI. Zhou \textit{et al.} \cite{zhou2019edge} surveyed the technologies of training and inference for deep learning models at network edges. Park \textit{et al.} \cite{park2019wireless} focused on the opportunities of utilizing edge AI for improving wireless communication, as well as realizing edge AI over wireless channels. Murshed \textit{et al.} \cite{murshed2019machine} mainly discussed different machine learning models and neural network models, different practical applications such as video analytics and image recognition, as well as various machine learning frameworks for enabling edge AI. Han \textit{et al.} \cite{deng2019edge} further considered  the convergence of edge computing and deep learning, i.e., the deep learning techniques for edge computing, as well as edge computing techniques for deep learning. 

Unlike existing survey papers \cite{zhou2019edge,park2019wireless,murshed2019machine,deng2019edge}, we shall present a comprehensive coverage to address the communication challenges for realizing AI at network edges. 
Edge AI is far from a trivial task of merely adopting the same computation and communication techniques in the cloud center. It requires learning performance aware joint design of computation and communication. Both distributed learning algorithms and distributed computing system architectures shall be customized according to the considered AI model, data availability, and the heterogeneous resources at edge nodes for reducing communication overheads during training and inference. We summarize the research topics on edge AI as algorithm-level designs and system-level designs, which are listed more specifically as follows: 
\begin{itemize}
    \item \textbf{Algorithm level:} At the algorithm level, the communication rounds of training a model can be reduced by accelerating convergence, while communication overhead per round can be reduced by information compression techniques (e.g., sparsification, quantization, etc.). We first survey different types of edge AI algorithms including the zeroth-order, first-order, second-order and federated optimization algorithm, as well as their applications in edge AI. For example, in the context of reinforcement learning, model-free based methods turn the reinforcement learning problem into zeroth-order optimization \cite{recht2019tour}. Although first-oder methods are widely used in DNNs training, second-order methods and federated optimization become appealing in edge AI given the growing computational capabilities of devices. 
    As we can see from Fig. \ref{fig:tradeoff}, the algorithm closer to the right side can potentially achieve better accuracy with less communication rounds, at the cost of more computation resources per round. Note that we list federated optimization methods separately due to its unique motivation to protect private data at each node. For each type of algorithms, there are a number of works focusing on further reducing communication cost.  We give a comprehensive survey on the algorithm level in Section III to address the communication challenges in edge AI. 
    \begin{figure}[h]
      \centering
      \includegraphics[width=0.9\linewidth]{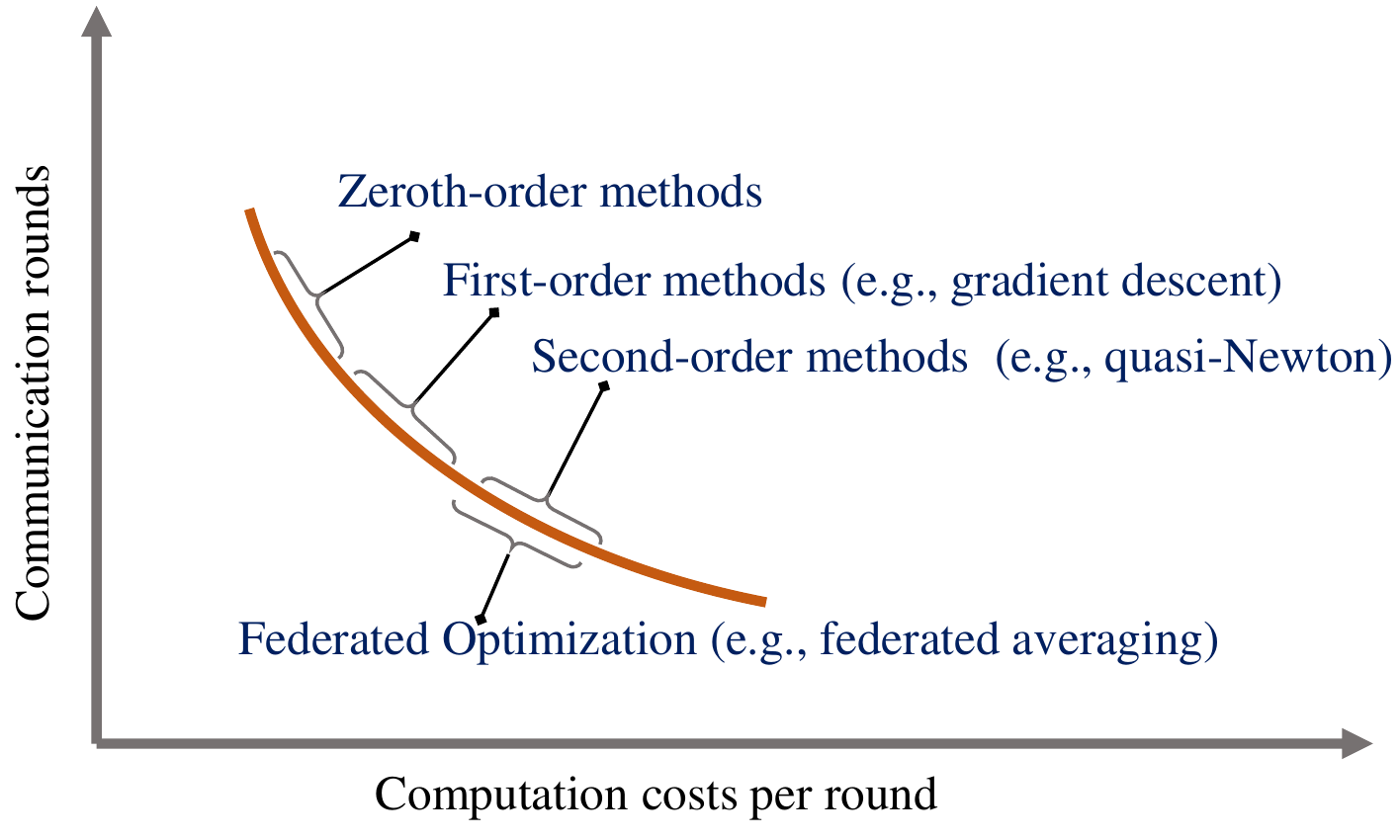}
      \caption{Communication rounds and computation costs per round for different types of training algorithms. The tradeoff has been widely used in designing algorithms, by performing more computation at each round in exchange for fewer number of communication rounds till convergence.}
      \label{fig:tradeoff}
    \end{figure} 

    \item \textbf{System level:} From the system perspective, data distribution (e.g., distributed across edge devices), model parameters (e.g., partitioned and deployed across edge devices and edge servers), computation (e.g., MapReduce), and communication mechanisms (e.g., aggregation at a central node) can be diverse in different applications. There are two main edge AI system architectures for training, i.e., the data partition system and model partition system, based on the availability of data and model. After training the AI model, model deployment is critical for achieving low-latency AI services. 
    There are also other general edge computing paradigms in edge AI systems that address the tradeoff between computation and communication via coding techniques. There are different types of communication problems arising from the deployment of machine learning algorithms on different system architectures, which typically involve the distributed mode and decentralized mode depending on the existence of a central node. We shall survey various system-level approaches to achieve efficient communications in Section IV. 
\end{itemize}
We summarize the main topics and highlighted technologies included in this paper in Table \ref{tab:topics}.

\begin{table*}
\centering
\caption{An Overview of Topics Covered in the Paper}
\label{tab:topics}
 \begin{tabular}{M{0.17\textwidth}M{0.22\textwidth}p{0.5\textwidth}}
 \toprule
 Category & Topic  &  {\qquad\qquad\qquad\qquad\qquad\qquad Representative Results} \\\toprule  
 \multirow{24}{0.17\textwidth}{\centering Communication-Efficient Algorithms for Edge AI} & \multirow{2}*{Zeroth-Order Methods} &  $\bullet$ Optimal rates for zeroth-order convex optimization \cite{Martin_TIT15zeroopt}  \\
&& $\bullet$ Distributed zeroth-order algorithms over time-varying networks \cite{yuan2014randomized,sahu2018distributed}
 \\\cmidrule(r){2-3}
 & \multirow{7}*{First-Order Methods} &  $\bullet$ Variance reduction for minimizing communication rounds \cite{yuan2018variance,lee2017distributed2} \\
 && $\bullet$ Gradient reuse for minimizing communication bandwidth \cite{chen2018lag,chen2018communication}\\
 && $\bullet$ Relating gradient quantization to limited feedback in wireless communication   \cite{du2019high}\\
 && $\bullet$ Communicating only important gradients for minimizing communication bandwidth \cite{chen2018adacomp,lin2018deep}\\\cmidrule(r){2-3}
 & \multirow{2}*{Second-Order Methods} & $\bullet$ Stochastic quasi-Newton methods \cite{schraudolph2007stochastic,byrd2016stochastic,moritz2016linearly} \\
 & & $\bullet$  Approximate Newton-type methods \cite{shamir2014communication,zhang2015disco,Mahoney_arXiv18,dunner2018distributed} \\\cmidrule(r){2-3}
 & \multirow{9}*{Federated Optimization} & $\bullet$ Federated averaging algorithm \cite{mcmahan2017communication} and dual coordinate ascent algorithm \cite{jaggi2014communication} for minimizing communication rounds \\
&& $\bullet$ Handling the system and statistical heterogeneity of distributed learning \cite{sahu2018fedprox,smith_nips2017federated} \\
& & $\bullet$ Compressing DNN models with vector quantization \cite{gong2014compressing}, binary weights \cite{courbariaux2015binaryconnect,courbariaux2016binarized,rastegari2016xnor}, randomized sketching \cite{chen2015compressing,Wenlin2016Compressing,Lin2019Towards}, network pruning \cite{lecun1990optimal,hassibi1993second,han2015learning,han2015deep,ullrich2017soft,louizos2017bayesian,Romberg_NIPS2017_6910,jiang2019layer}, sparse regularization \cite{lebedev2016fast,wen2016learning,oymak2018learning}, and structural matrix designing for minimizing communication bandwidth  \cite{sainath2013low,denton2014exploiting,jaderberg2014speeding,denil2013predicting,rigamonti2013learning,Amos2015Learning,sindhwani2015structured,cheng2015exploration,yang2015deep}
 \\\midrule
 \multirow{27}{0.17\textwidth}{\centering Communication-Efficient Edge AI Systems} & \multirow{7}{0.22\textwidth}{\centering Data Partition Based Edge Training Systems} & $\bullet$ Fast aggregation via over-the-air computation \cite{yang2018federated,zhu2018low,amiri2019machine,amiri2019federated}\\
 && $\bullet$  Aggregation frequency control with limited bandwidth and computation resources \cite{Kevin_arXiv18federated,zhou2018convergence,wang2018adaptive}\\
 && $\bullet$  Data reshuffling via index coding and pliable index coding for improving training performance \cite{Ramchandran_DML18,song2017pliable,jiang2019pliable}\\
 && $\bullet$  Straggler mitigation via coded computing \cite{Tandon_pmlr-v70-tandon17a,ye2018communication,raviv2018gradient,halbawi2018improving,li2018near,ozfaturay2018speeding,bitar2019stochastic,maity2019robust,horii2019distributed}\\
 && $\bullet$  Training in decentralized system mode \cite{pramod2018elastic,daily2018gossipgrad,blot2019distributed,blot2016gossip,nedic2018network,adjodah2019communication,neglia2019role,patarasuk2009bandwidth,jin2016scale,sergeev2018horovod,reisizadeh2019codedreduce} \\\cmidrule(r){2-3}
& \multirow{7}{0.22\textwidth}{\centering Model Partition Based Edge Training Systems} & $\bullet$ Model partition across a large number of nodes to balance computation and communication \cite{Dean_ICML17placement,harlap2018pipedream,huang2019gpipe}\\
 &&$\bullet$ Model partition across edge device and edge server to avoid the exposure of users’ data \cite{mao2018privacy,wang2018not}\\
 &&$\bullet$ Vertical architecture for privacy with vertically partitioned data and model \cite{yang2019federated,vaidya2003privacy,kantarcioglu2004privacy,gascon2016secure,yu2006privacy_svm,vaidya2005privacy,hardy2017private}  \\\cmidrule(r){2-3}
&  \multirow{10}{0.22\textwidth}{\centering Computation Offloading Based Edge Inference Systems} & Server-based edge inference:
\begin{itemize}
    \item Partial data transmission for communication-efficient inference \cite{ding2019communication,paull2015communication,mohanarajah2015cloud,chen2015glimpse}
    \item Raw data encoding for communication-efficient inference \cite{liu2018deepn,chinchali2018neural}
    \item Cooperative downlink transmission for communication-efficient inference \cite{yang2019energy,hua2019reconfigurable}
\end{itemize}\vspace{-1em} \\
&& Device-edge joint inference:
\begin{itemize}
    \item Early exit: \cite{teerapittayanon2016branchynet,teerapittayanon2017distributed}
    \item Encoded transmission and pruning for compressing the transmitted data \cite{ko2018edge,shi2019improving} 
    \item Coded computing for cooperative edge inference \cite{zhang2019model}
\end{itemize}\vspace{-1em} \\\cmidrule(r){2-3}
 &  \multirow{2}{0.22\textwidth}{\centering General Edge Computing Systems} & $\bullet$ Coding techniques for efficient data shuffling \cite{Ali_ComMag17,Ali_CCT16,Ali_arXiv16WDC,Yuanming_WDCTSP18,li2018wireless,ha2018wireless,ji2018fundamental,parrinello2018coded,prakash2018coded} \\
 && $\bullet$ Coding techniques for straggler mitigation \cite{parrinello2018coded,reisizadeh2017latency,zhao2019node,kosaian2018learning} 
  \\\bottomrule
    \end{tabular}
\end{table*}

\section{Communication-efficient Algorithms for Edge AI}
Distributed machine learning has been mainly investigated in the environment with abundant computing resources, large memory, and high-bandwidth networking, e.g., in cloud data centers. The extension to the edge AI system is highly non-trivial due to the isolated data at distributed mobile devices, limited computing resources, and the heterogeneity in communication links. Communication-efficient methods will be critical to exploit the distributed data samples and utilize various available computing resources for achieving excellent learning performance.  This section introduces communication-efficient approaches for edge AI at the algorithmic level, including zeroth-order methods, first-order methods, second-order methods, and federated optimization. As illustrated in Fig. \ref{fig:tradeoff}, these methods achieve different tradeoffs among the local computation and communication cost. Fig. \ref{fig:algorithms} provides illustrations of local operations and communication messages of different methods.

\subsection{Communication-Efficient Zeroth-Order Methods}
\begin{figure*}[htb]
  \centering
  \subfigure[Zeroth-order method]{\includegraphics[width=\columnwidth]{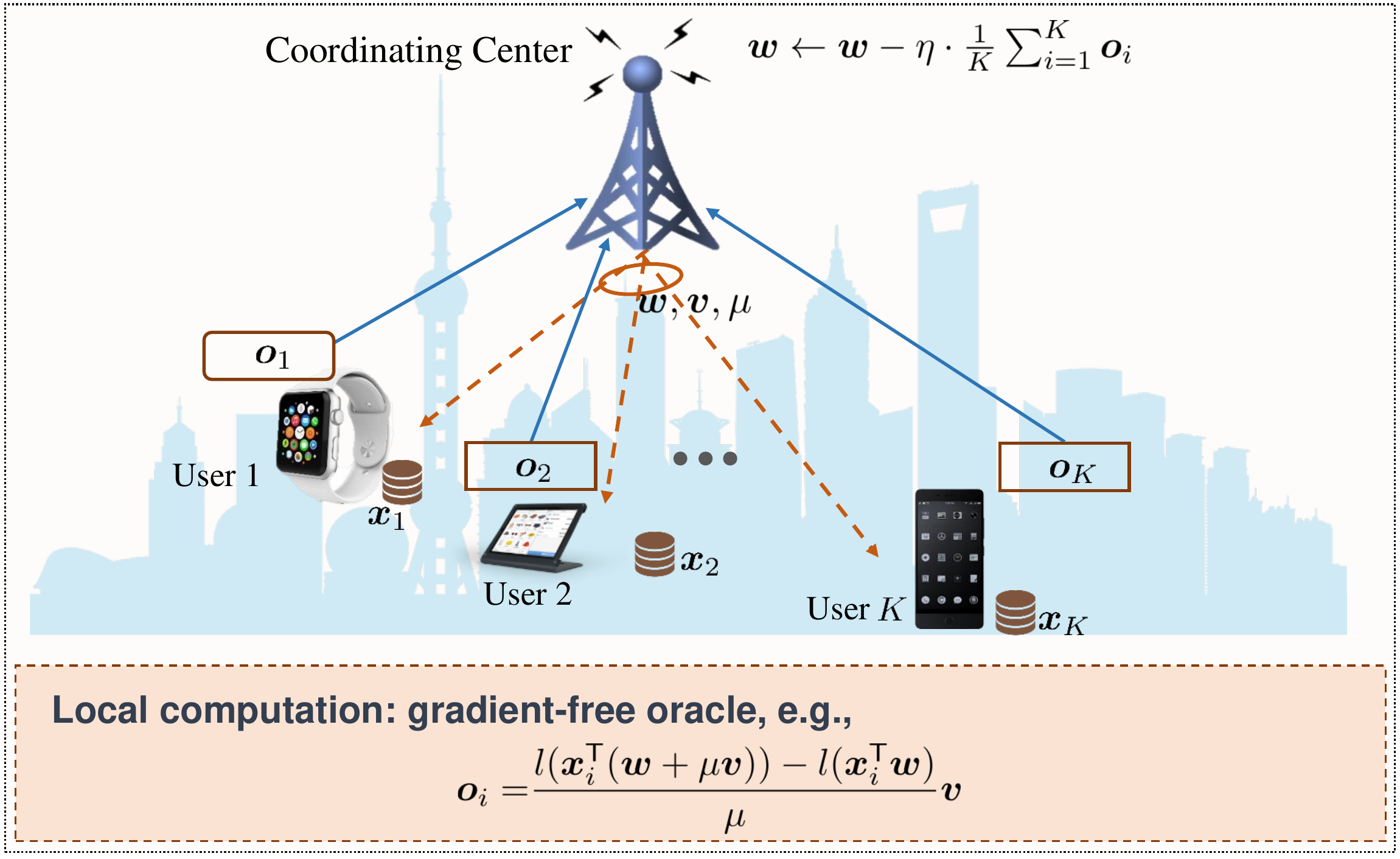}\label{fig:zeroth_order}}
  \subfigure[First-order method]{\includegraphics[width=\columnwidth]{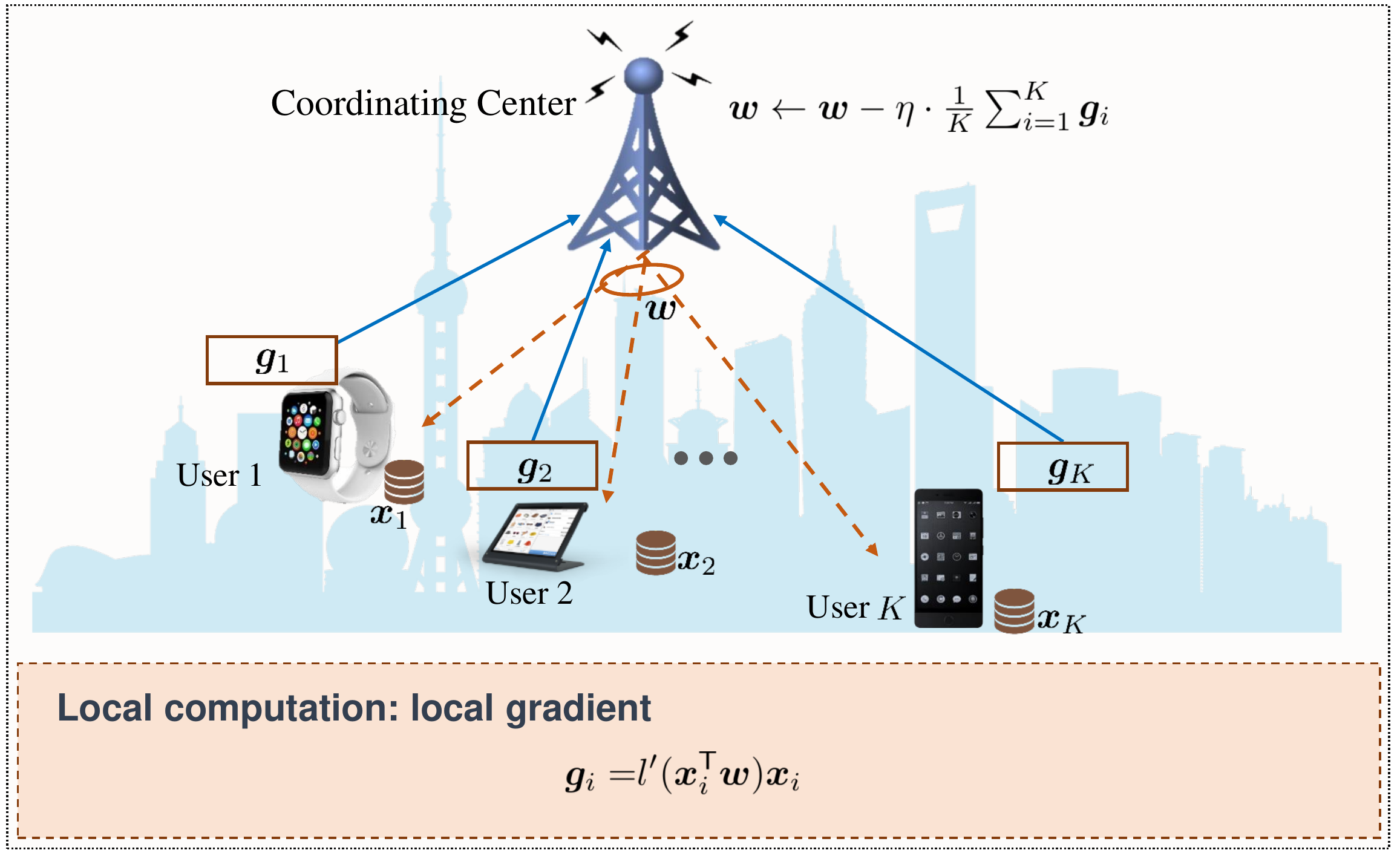}\label{fig:first_order}}
  \subfigure[Second-order method]{\includegraphics[width=\columnwidth]{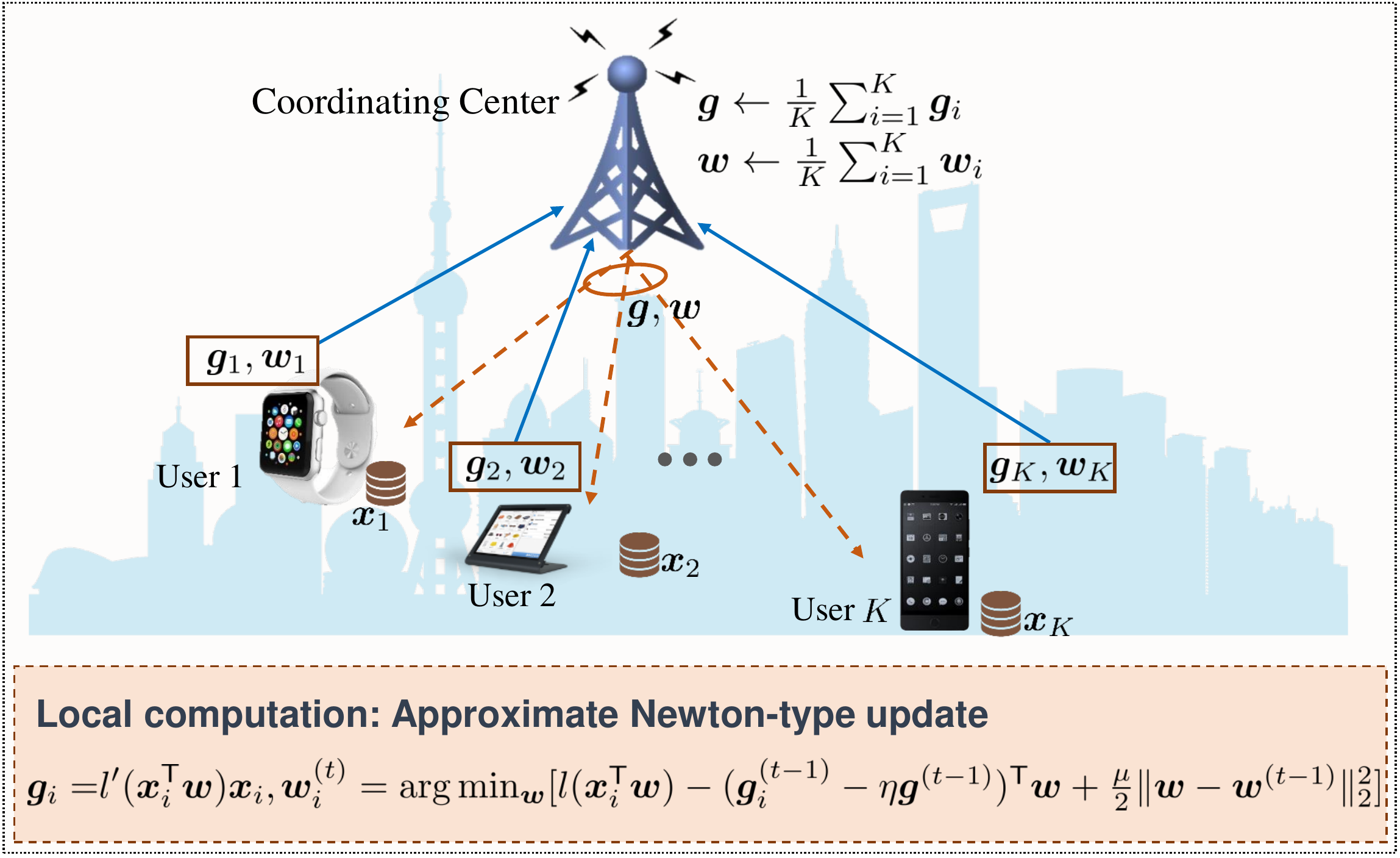}\label{fig:second_order}}
  \subfigure[Federated Optimization]{\includegraphics[width=\columnwidth]{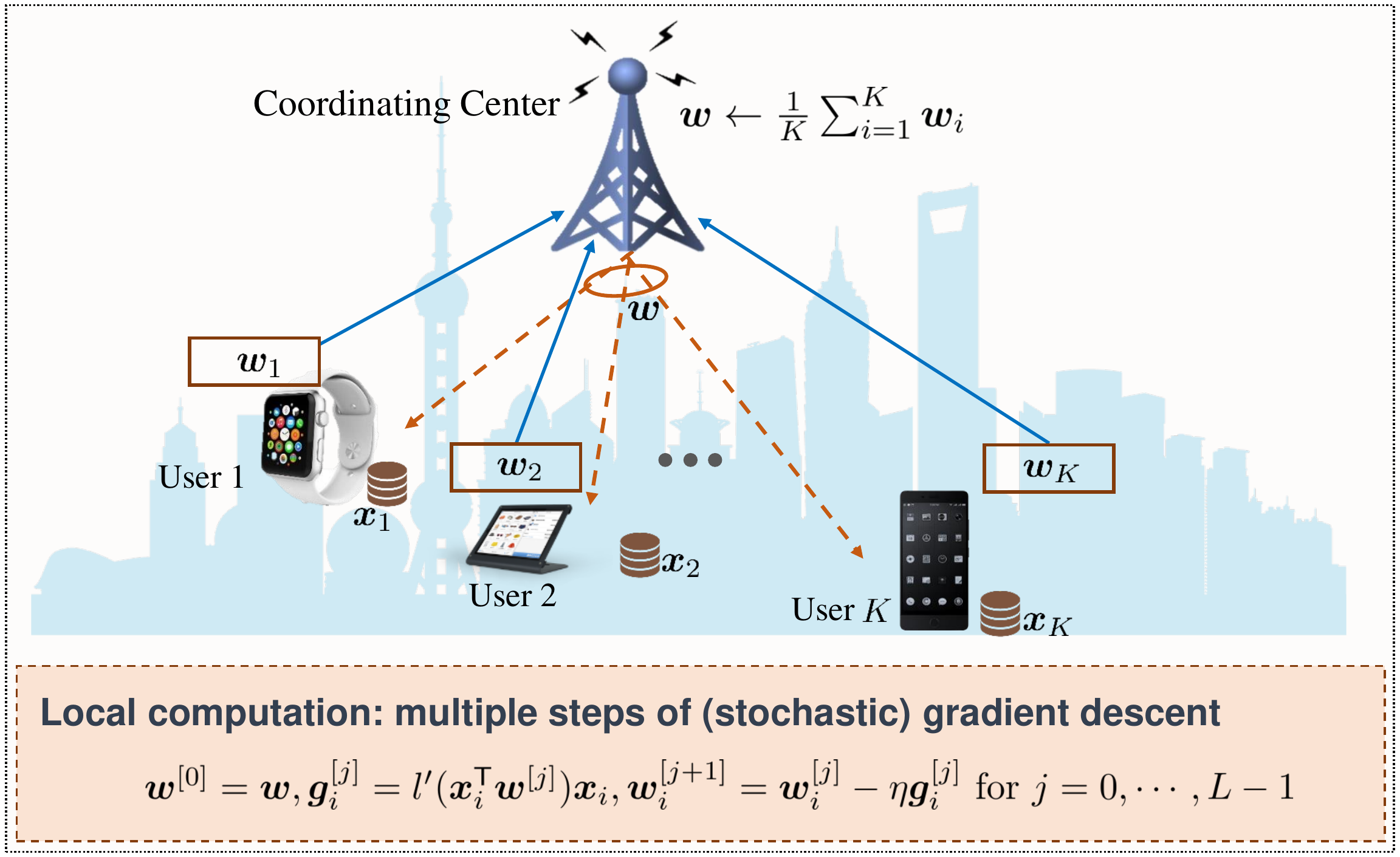}\label{fig:federated}}
  \caption{Illustration of different optimization methods for model training. As a typical example, a generalized linear model is trained where each node $i$ has one data instance $\bm{x}_i$. That is, the target is to optimize $\min_{\bm{w}}\frac{1}{K}\sum_{i=1}^{K}l(\bm{w}^{\sf T}\bm{x}_i)$, where $l$ is the loss function. The first-order derivative of function $\ell$ is denoted as $\ell^\prime$. a) Zeroth-order method: only the function value can be evaluated during training \cite{nesterov2017random}. b) First-order method: gradient descent. c) Second-order method: DANE \cite{shamir2014communication}. d) Federated optimization: federated averaging algorithm \cite{mcmahan2017communication}.}
  \label{fig:algorithms}
\end{figure*}

Zeroth-oder (derivative-free) methods \cite{nesterov2017random} are increasingly adopted in the applications where only the function value is available, but the derivative information is computationally difficult to obtain, or is even not well defined. For the distributed setting with a central coordinating center shown in Fig. \ref{fig:zeroth_order}, only a function value scalar is required to be transmitted to the central node in uplink transmission. In the field of reinforcement learning, zeroth-order methods have been widely used for policy function learning without ever building a model \cite{recht2019tour}. Zeroth-order methods have also been adopted to black-box adversarial attacks on deep neural networks (DNNs) since most real world systems do not release their internal DNNs structure and weights \cite{chen2017zoo}.

In zeroth-order optimization algorithms, the full gradients are typically  estimated via gradient estimators based on only the function values \cite{conn2009introduction}. For instance, we can use the quantity $(l(\bm{w}+\mu\bm{v})-l(\bm{w}))\bm{v}/\mu$ to approximate the gradient of function $l(\bm{w})$ at point $\bm{w}$. It was shown in \cite{Martin_TIT15zeroopt} that this kind of derivative-free algorithm only suffers a factor of at most \(\sqrt{d}\) in the convergence rate over traditional stochastic gradient methods for \(d-\)dimensional convex optimization problems. Under time-varying random network topologies, recent studies \cite{yuan2014randomized,sahu2018distributed} have investigated the distributed zeroth-order optimization algorithms for unconstrained convex optimization in multi-agent systems. Convex optimization with a set of convex constraints have been studied in \cite{yuan2015zeroth,pang2018exact}. Nonconvex multi-agent optimization  has been studied in \cite{hajinezhad2019zone} for  different types of network topologies, including  undirected connected networks or star networks under the setting where the agent can only access the values of its local function.  

To develop communication-efficient distributed zeroth-order optimization methods, there have been a number of works on reducing the number of per-device communication. For instance, it was proposed in \cite{sahu2018non} that at each iteration each device communicates with its neighbors with some probability that is independent from others and the past, and this probability parameter decays to zero at a carefully tuned rate. For such a distributed zeroth-oder method, the convergence rate of the mean squared error of solutions is established in terms of the communication costs, i.e., the number of per-node transmissions to neighboring nodes in the network, instead of the iteration number. The subsequent work \cite{sahu2018communication} improved the convergence rate under additional smoothness assumptions. Quantization techniques are also adopted to reduce the communication cost per communication round.  The paper \cite{ding2017distributed} considered a distributed  gradient-free algorithm for multi-agent convex optimization, where the agents can only exchange quantized data information due to limited bandwidth. In the extreme case considered in \cite{liu2018signsgd}, each estimated gradient is further quantized into 1 bit, which enjoys high communication efficiency in distributed optimization scenarios.

\subsection{Communication-Efficient First-Order Methods}
First-order optimization methods are the most commonly used algorithms in machine learning, which are mainly based on gradient descent methods as shown in Fig. \ref{fig:first_order}. The idea of gradient descent methods is to iteratively update variables in the opposite direction of the gradients of the loss function at that point with an appropriate step size (a.k.a., a learning rate).  As the computational complexity at each iteration scales with the number of data samples and the dimension of the model parameter, it is generally infeasible to train large machine learning models with tremendous amount of training data samples on a single device. Therefore,  distributed training techniques have been proposed to mitigate the computation cost, with additional communication costs. Meanwhile, as the training dataset becomes larger and larger, stochastic gradient descent (SGD) method emerges as an appealing solution, in which only one training sample is used to compute the gradient at each iteration. In edge AI systems with inherently isolated data,  distributed realizations of SGD will play a key role and should be carefully investigated. 

To apply first-order methods in large-scale distributed edge AI systems,  the substantial demand for communication among devices for gradient exchange is one of the main bottlenecks.  One way to address this issue is to reduce the communication round by accelerating the convergence rate of the learning algorithms. Another approach is to reduce the communication overhead per round, which includes gradient reuse method, quantization, sparsification, and sketching based compression methods. These two approaches are elaborated in the following.

\subsubsection{Minimizing Communication Round}
We first consider an extreme case.
The distributed optimization approach with the minimum communication round, i.e., only one communication round, is that each device performs independent optimization. For example, each node adopts SGD to compute local model parameters, and a server then averages these model parameters in the end. As shown in \cite{zinkevich2010parallelized}, the overall run time decreases significantly as the number of devices increases for some learning tasks. Subsequently, it was shown in \cite{zhang2013communication} that this one-round communication approach can achieve the same order-optimal sample complexity in terms of mean-squared error of model parameters as the centralized setting under a reasonable set of conditions.   The order-optimal sample complexity can be obtained by performing a stochastic gradient-based methods on each devices \cite{zhang2013communication}. However, one round communication restricts the ability to exchange information during training, which is in general not sufficient for training large models (e.g., DNNs) to achieve the target accuracy in practice. 

In general settings where devices upload their local gradients to a fusion center at each iteration, it is critical to reduce the communication round by accelerating the convergence rate of the algorithm. Shamir and Srebro \cite{shamir2014distributed} proposed to  accelerate mini-batch SGD by using the largest possible mini-batch size that does not hurt the sample complexity, and it shows that the communication cost decreases linearly with the size of the mini-batch. Yuan \textit{et al.} \cite{yuan2018variance} proposed an amortized variance-reduced gradient algorithm for a decentralized setting, where each device collects data that is spatially distributed and all devices are only allowed to communicate with direct neighbors. In addition, a mini-batch strategy is adopted by \cite{yuan2018variance} to achieve communication efficiency. 
However, it has been shown in \cite{keskar2016large,yin2018gradient} that too large mini-batch sizes will result in a degradation in the generalization of the model. In practice, additional efforts should be taken to reduce this generalization drop. For instance, it was shown in \cite{goyal2017accurate} that training with large minibatch sizes up to \(8192\) images achieves the same accuracy as small mini-batch settings  by adjusting learning rates as a function of mini-batch size. This idea was also adopted by \cite{zhang2019distributed}  to train DNNs for automatic speech recognition tasks  using large batch sizes in oder to accelerate the total training process.

The statistical heterogeneity of data hinders the fast convergence of first-order algorithms. To address this issue, there have been lots of efforts. Arjevani and Shamir \cite{arjevani2015communication} studied the scenarios where each device has access to a different subset of data to minimize the averaged loss function over all devices. They established a lower bound on the rounds of communication, which is shown to be achieved by the algorithm of \cite{zhang2018communication} for quadratic and strongly convex functions. But how to design optimal algorithms in terms of communication efficiency for general functions remains an open problem.  By utilizing additional storage space of devices, Lee \textit{et al.} \cite{lee2017distributed2} proposed to assign two subsets of data to each device. The first subset is from a random partition and the second subset is randomly sampled with replacement from the overall datasets. Since each device has access to both data subsets, the authors proposed a distributed stochastic variance reduced gradient method to minimize the communication round, in which the batch gradients are computed in parallel on different devices and the algorithm utilizes the local data sampled with replacement to construct the unbiased stochastic gradient in each iterative update.
For non-convex optimization problems, Garber \textit{et al.} \cite{garber2017communication} proposed a stochastic distributed algorithm to solve the principal component analysis problem, which  gives considerable acceleration in terms of communication rounds over previous distributed algorithms.

\subsubsection{Minimizing Communication Bandwidth}
Another series of works focus on reducing the size of local updates from each device, thereby reducing the overall communication cost. In the following, we review three representative techniques, i.e., gradient reuse, gradient quantization, and gradient sparsification.

\begin{itemize}[leftmargin=0pt,itemindent=1.5em,align=left,topsep=0.5em,itemsep=0.5em]\setlength{\parindent}{1em}
    \item \textbf{Gradient reuse:}
To minimize a sum of smooth loss functions distributed among multiple devices, considering that the gradients of some devices vary slowly between two consecutive communication rounds, a lazily aggregated gradient (LAG) method was proposed by \cite{chen2018lag} which uses outdated gradients of these devices at the fusion center. Specifically, theses devices upload nothing during this communication round, which is able to reduce communication overheads per round significantly.  Theoretically, it was shown in \cite{chen2018lag} that LAG achieves the same order of convergence rates as the batch gradient descent method under the cases where the loss functions are strongly-convex, convex, or nonconvex smooth. If the distributed datasets are heterogeneous, LAG can achieve a target accuracy with considerably less communication costs measured as the total number of transmissions over all the devices in comparison with the batch gradient descent method.  In addition, a similar gradient reuse idea was adopted in distributed reinforcement learning to achieve  communication efficient training \cite{chen2018communication}.

\item \textbf{Gradient quantization:}
To reduce the communication cost of gradient aggregation, some scalar quantization methods have been proposed to compress the gradients by a small number of bits instead of using floating-point representation. To estimate the mean of the gradient vectors collected from devices, Suresh \textit{et al.} \cite{suresh2017distributed} analyzed the mean squared error for several quantization schemes without probabilistic assumptions on the data from the information theoretic perspective. In the view of distributed learning algorithms. Alistarh \textit{et al.} \cite{alistarh2017qsgd} has investigated the quantized stochastic gradient descent (QSGD) to study the trade-off between communication costs and convergence guarantees. Specifically, each device can adjust the number of bits sent per iteration according to the variance added to the device. As demonstrated in \cite{alistarh2017qsgd}, each device can transmit no more than \(2.8n+32\) bits per iteration in expectation, where \(n\) is the number of model parameters, while the variance is only increased by a factor of \(2\).  Compared to full precision SGD,  using approximately \(2.8n\)  bits of communication per iteration as opposed to \(32n\) bits  will only result in  at most \(2\times\) more iterations, which leads to bandwidth savings of approximately \(5.7\times\).
For distributed training the shallowest neural networks consisting of a single rectified linear unit, it was shown in \cite{mousavi2019fitting} that the quantized stochastic gradient method converges to the global optima at a linear convergence rate. Seide \textit{et al.} \cite{seide20141} proposed to quantize the gradient 
using only one bit, achieving a 10 times speed-up on distributed training of speech DNNs with a small accuracy loss.
Theoretically, Bernstein \textit{et al.} \cite{bernstein2018signsgd} provided rigorous  analysis for the sign-based distributed stochastic gradient descent algorithm, where each device sends the sign information of the gradients to a fusion center and the sign information of the aggregated gradients signs is returned to each device for updating model parameters. This scheme is shown to achieve the same reduction in variance as full precision distributed SGD and converge to a stationary point of a general non-convex function.

As pointed out in \cite{tang2018communication,yu2018gradiveq}, scalar quantization methods fail under decentralized networks without a central aggregation node. To address this issue, extrapolation compression and difference compression methods were proposed in \cite{tang2018communication}, and a gradient vector quantization technique was proposed in \cite{yu2018gradiveq} to exploit the correlations between CNN gradients. Vector quantization \cite{gersho2012vector} by jointly quantizing all entries of a vector can achieve the optimal rate-distortion trade-off, which, however, comes at the price of high complexity that increases with the vector length. Interestingly, it was found in \cite{du2019high} that 
Grassmannian quantization, a vector quantization method that has already been widely adopted in wireless communication for limited feedback, can be applied for gradient quantization. Limited feedback is an area of studying efficient feedback of quantized vectors from a receiver to a transmitter for channel adaptive transmission schemes since the communication cost of feedback is extremely high in massive MIMO communication systems. This motivated \cite{du2019high} to develop an efficient Grassmannian quantization scheme for high-dimensional gradient compression in distributed learning.

Additionally, Jiang \textit{et al.} \cite{jiang2018sketchml} proposed to use quantile sketch, a non-uniform quantization method for gradient compression. Sketch is a technique of approximating input data with a probablistic data structure. In \cite{jiang2018sketchml}, the gradient values are summarized into a number of buckets, whose indices are further encoded by a binary representation since the number of buckets is relatively small.

\item \textbf{Gradient sparsification:}
The basic idea behind gradient sparsification is to communicate only important gradients according to some criteria. This is based on the observation that many gradients are normally very small during training. Strom \cite{strom2015scalable} proposed to leave out the gradients below a predefined constant threshold. Chen \textit{et al.} \cite{chen2018adacomp} proposed AdaComp via localized selection of gradient residues, which automatically
tunes the compression rate depending on local activity. It was demonstrated that  AdaComp can achieve a compression ratio of around \(200\times\) for fully-connected layers and \(40\times\) for convolutional layers without noticeable degradation of top-1 accuracy on ImageNet dataset.
Deep gradient compression was proposed in \cite{lin2018deep} based on a gradient sparsification approach, where only gradients exceeding a threshold are communicated, while the remaining  gradients are accumulated until they reach the threshold. Several techniques including  momentum correction, local gradient clipping, momentum factor
masking, and warm-up training are adopted to preserve the accuracy. This deep gradient compression approach is shown to achieve a gradient compression ratio from \(270\times\) to \(600\times\) without losing accuracy for a wide range of DNNs and RNNs \cite{lin2018deep}. 
In \cite{wangni2018gradient}, to ensure the sparsified gradient to be unbiased, the authors proposed to drop some coordinates of the stochastic gradient vectors randomly and amplify the rest of the coordinates appropriately. 
For both convex and non-convex smooth objectives, under analytic assumptions, it was shown in \cite{alistarh2018convergence} that sparsifying gradients by magnitude with local error correction provides convergence guarantees. Thus, providing a theoretical foundation for numerous empirical results on training large-scale  recurrent neural networks on a wide range of applications.

\end{itemize}

\subsection{Communication-Efficient Second-order Methods}
First-order algorithms only require the computation of gradient-type updates, and thus reduce the amount of local computation at each device. But the main drawback is that the required number of communication rounds is still huge due to the slow convergence rate. It thus motivates to exploit second-order curvature information into distributed learning algorithms to improve the convergence rate for edge training. However, exact second-order methods require the computation, storage and even communication of a Hessian matrix, which results in tremendous overhead. Therefore, one has to resort to approximate methods such as illustrated in Fig. \ref{fig:second_order} \cite{shamir2014communication}. The works on communication-efficient second-order methods can be categorized into two types. One is to maintain a global approximated inverse Hessian matrix in the central node, and the other line of works propose to solve a second-order approximation problem locally at each device.



A common approach to develop approximate second-order methods is to take the merits of the well-known quasi-Newton method, namely Limited-memory Broyden Fletcher Goldfarb Shanno (L-BFGS) \cite{liu1989limited}, which avoids the high cost of computing the inversion of Hessian matrix via directly estimating the inverse Hessian matrix. In learning with large amounts of training data, it is a critical problem to develop a mini-batch stochastic quasi-Newton method. However, directly extending L-BFGS to a stochastic version does not result in a stable approximation of the inverse Hessian matrix. Schraudolph \textit{et al.} \cite{schraudolph2007stochastic} developed a stochastic L-BFGS for online convex optimization without line search, which is often problematic in a stochastic algorithm. But there may be a high level of noise in its Hessian approximation. To provide stable and productive Hessian approximations, Byrd \textit{et al.} \cite{byrd2016stochastic} developed a stochastic quasi-Newton method by updating the estimated inverse Hessian matrix every $L$ iterations using sub-sampled Hessian-vector products. The inverse Hessian matrix maintained in a central node is updated by collecting only a Hessian-vector product update at each device. Moritz \textit{et al.} \cite{moritz2016linearly} proposed a linearly convergent stochastic L-BFGS algorithm via obtaining a more stable and higher precision estimation of the inverse Hessian matrix, but it requires higher computation and communication overhead at each round.

Another main idea of communication-efficient second-order methods is to solve a second-order approximation problem at each device without maintaining and computing a global Hessian matrix. To reduce the communication overhead at each round, Shamir \textit{et al.} \cite{shamir2014communication} proposed a distributed approximate Newton-type method named as ``DANE'' by solving an approximate local Newton system at each device with a global aggregation step, which only requires the same communication bandwidth as first-order distributed learning algorithms. Subsequently, the algorithm ``DiSCO'' proposed in \cite{zhang2015disco} solved a more accurate second-order approximation at per communication round by approximately solving the global Newton system with a distributed preconditioned conjugate gradient method. It reduces the communication rounds compared with ``DANE'', while the computation cost at the master machine grows roughly cubically with the model dimension. Wang \textit{et al.}
\cite{Mahoney_arXiv18} proposed an improved approximate Newton method ``GIANT'' to further reduce the communication round via conjugate gradient steps at each device, which is shown to  outperform ``DANE'' and ``DiSCO''. Note that the communication of these approaches involves the transmission of a global update to each device and the aggregation of local update from each device at per round, both with the same size as the number of model parameters. However, the convergence results of ``DANE'', ``DiSCO'', and ``GIANT'' require a high accuracy solution to the subproblem at each device. An adaptive distributed Newton method was proposed in \cite{dunner2018distributed} by additionally transmitting a scalar parameter accounting for the information loss of distributed second-order approximation at per round, which outperforms ``GIANT'' in numerical experiments. 

\subsection{Communication-Efficient Federated Optimization}
In the edge training system, the local dataset at each device is usually only a small subset of the overall dataset. Furthermore, the rapid advancement of CPUs and GPUs on mobile devices makes the computation essentially free in comparison to the communication cost. Thus, a natural idea is to use additional local computation to decrease the communication cost.
Federated optimization \cite{konevcny2015federated} is a framework of iteratively performing a local training algorithm (such as multiple steps of SGD as illustrated in Fig. \ref{fig:federated}) based on the dataset at each device and aggregating the local updated models, i.e., computing the average (or weighted average) of the local updated model parameters. This framework provides additional privacy protection for data, and has the potential of reducing the number of communication rounds for aggregating updates from a large number of mobile devices. The concern of data privacy and security is becoming a worldwide major issue, especially for emerging high-stake applications in intelligent IoT, which prompted governments to enact new regualtions such as General Data Protection Regulation (GDPR) \cite{euGDPR2016}. 
There are a line of works studying federated optimization algorithms \cite{jaggi2014communication,mcmahan2017communication,smith_nips2017federated} to reduce the communication rounds. In addition, a number of model compression methods have been proposed to reduce the model size, either during the local training process or compressing the model parameters after local training, which can further reduce the communication cost of aggregation for federated optimization \cite{konevcny2016federated}. These methods are reviewed in this part.

\subsubsection{Minimizing Communication Round}
Jaggi et cl. \cite{jaggi2014communication} proposed a framework named ``CoCoA'' by leveraging the primal-dual structure of a convex loss function of general linear models with a convex regularization term. In each communication round, each mobile device performs multiple steps of a dual optimization method based on local dataset in exchange for fewer communication rounds, followed by computing the average of updated local models. Motivated by \cite{jaggi2014communication}, authors in \cite{smith_nips2017federated} further proposed a  communication-efficient federated optimization algorithm called ``MOCHA'' for multi-task learning. By returning an additional accuracy level parameter, it is also capable of dealing with straggling devices. However, these algorithms are not suitable for a general machine learning problem when the strong duality fails or the dual problem is difficult to obtain.

The Federated Averaging (FedAvg) \cite{mcmahan2017communication} algorithm is another communication-efficient federated optimization algorithm by updating local model at each device with a given number of SGD iterations and model averaging. It is generalized from the traditional one-shot averaging algorithm \cite{zhang2012communication} that is applicable only when the data samples at each device are drawn from the same distribution. In per round of communication, each device performs a given number of steps of SGD with a global model as the initial point, and the aggregated global model is given by the weighted average of all local models. The weights are chosen as the sizes of the local training dataset, which is shown to be robust to not independently and identically distributed (non-IID) data distribution and unbalanced data across mobile devices. Wang and  Joshi \cite{wang2018cooperative} provided the convergence result of the FedAvg algorithm to a stationary point. To reduce the costly communication with the remote cloud, edge server assisted hierarchical federated averaging was proposed in \cite{liu2020client}. By exploiting the highly efficient local communications with edge servers, it achieves significant training speedup compared with the cloud-based approach. With infrequent model aggregation at the cloud, it also achieves higher model performance than edge-based training, as data from more users can be accessed.


For the FedAvg algorithm, the steps of local SGD at each device should be chosen carefully given the existence of statistical heterogeneity, i.e., when the local data across devices are non-IID. If too many steps of SGD are performed locally, the learning performance will be degraded. To address this problem, the FedProx algorithm \cite{sahu2018fedprox} was proposed by adding a proximal term in the local objective function to restrict the local updated model to be close to the global model, instead of initializing each local model with the global updated at each communication round. Its convergence guarantees are also provided via characterizing the heterogeneity with a device dissimilarity assumption.
Numerical results demonstrate that FedProx is more robust to the statistical heterogeneity across devices. 

\subsubsection{Minimizing Communication Bandwidth}
Transmitting the model parameters per communication round  generally results in a huge communication overhead since the number of model parameters can be very large. Therefore, it is important to reduce the model size to alleviate the communication overhead \cite{konevcny2016federated}. To this end, model compression is one of the promising approaches. We survey the main techniques adopted in model compression in this subsection.

\begin{itemize}[leftmargin=0pt,itemindent=1.5em,align=left,topsep=0.5em,itemsep=0.5em]\setlength{\parindent}{1em}
    \item \textbf{Quantization:}
Quantization compresses DNNs by representing the weights by fewer bits instead of adopting the 32-bit floating point format. The works \cite{gong2014compressing,wu2016quantized} adopt $k$-means clustering to the weights of a pre-trained DNN. At the training stage, it has been shown that DNNs can be trained using only 16-bit wide fixed-point number representation by stochastic rounding \cite{gupta2015deep}, which induces little to no degradation in the classification accuracy. 
In the extreme case, the weights are represented by $1$-bit, but the naive approach that binarizes pre-trained DNNs directly shall bring performance loss significantly. Therefore, the main idea behind binarization is to learn the binary weights or activation during training, which are thoroughly investigated in \cite{courbariaux2015binaryconnect,courbariaux2016binarized,rastegari2016xnor}. This kind of method allows a substantial computational speedup on devices due to the bit-wise operations. It may also reduce the communication cost in federated learning significantly as the weights are represented by $1$-bit.

\item 
\textbf{Sketching:}
Randomized sketching \cite{martin_2016iterative,choi2019large} is a powerful tool for dimensionality reduction, which can be applied to model compression.
In \cite{chen2015compressing}, HashedNet sketches the weights of neural networks using a hash function, and enforces all weights that are mapped to the same hash bucket to share a single parameter value. But it is only applicable to fully connected neural networks. The subsequent work \cite{Wenlin2016Compressing} extended it to CNNs, which is achieved by first converting filter weights to the frequency domain and then grouping the corresponding frequency parameters into hash buckets using a low-cost hash function. Theoretically, it was shown in \cite{Lin2019Towards} that such Hashing-based neural networks have nice properties, i.e., local strong convexity and smoothness around the global minimizer.

\item 
\textbf{Pruning:}\label{subsec:pruning}
Network pruning generally compresses DNNs by removing the connections, filters or channels according to some criteria. Early works include the Optimal Brain Damage \cite{lecun1990optimal} and the Optimal Brain Surgeon \cite{hassibi1993second}, which proposed to remove the connections between neurons based on the  Hessian of the loss function given a trained DNN. Recently, a line of research is to prune redundant, less important connections in a pre-trained DNN. For instance,
the work in \cite{han2015learning} proposed to prune the unimportant weights of a pre-trained network and  retrain the network to fine tune the weights of the remaining connections, which reduces the number of parameters of AlexNet by $ 9\times $ without harming the accuracy. Deep compression was proposed in \cite{han2015deep} to compress DNNs via three stages, i.e., pruning, trained quantization and Huffman coding, which yields considerably compact DNNs. For example,  the storage size of AlexNet is reduced by $35\times$ on the ImageNet dataset without loss of accuracy. From a Bayesian point of view, network pruning was also investigated in \cite{ullrich2017soft,louizos2017bayesian}.
However, such heuristic methods present no convergence guarantees. Instead, Aghasi \textit{et al.} \cite{Romberg_NIPS2017_6910} proposed to prune the network layer-by-layer via convex programming, which also shows that the overall performance drop can be bounded by the sum of the reconstruction error of each layer. Subsequently, iterative reweighed optimization has been adopted to further prune the DNN with convergence guarantees \cite{jiang2019layer}.

\item 
\textbf{Sparse regularization:}
There is a growing interest in learning compact DNNs without pre-training, which is  achieved by adding regularizers to the loss function during training in order to induce sparsity in DNNs. In \cite{lebedev2016fast}, the authors proposed to use a regularizer based on $ \ell_{2,1} $-norm to induce group-sparse structured convolution kernels when training CNNs, which leads to computational speedups.  To remove trivial filters, channels and even layers at the training stage, the work in \cite{wen2016learning} proposed to add structured sparsity regularization on each layer. Theoretically, the convergence behavior of gradient descent algorithms for learning shallow compact neural networks was depicted in \cite{oymak2018learning}, which also shows the  required sample complexity for efficient learning.

\item 
\textbf{Structural matrix designing:}
The main idea behind low-rank matrix factorization approaches for compressing DNNs is to apply low-rank matrix factorization techniques to the weight matrix of DNNs. For a low rank matrix $\bm A\in\mathbb{R}^{m\times n}$ with $\rank{(\bm A)}=r$, we can represent it as $\bm A=\bm B\bm C$ where $\bm B\in\mathbb{R}^{m\times r}$. Therefore, we reduce the total parameters from $mn$ to $mr+nr$, which is able to reduce the computational complexity and storage. For example, the work in \cite{sainath2013low} showed that the number of parameters of the DNNs can be reduced by $30\%-50\%$ for large vocabulary continuous speech recognition tasks via low-rank matrix factorization of the final weight layer. In \cite{denton2014exploiting}, in order to accelerate convolution, each convolutional layer is approximated by a low-rank matrix, and different approximation metrics are studied to improve the  performance. The work in \cite{jaderberg2014speeding} proposed to speed up the convolutional layers by constructing a low rank basis of rank-one filters for a pre-trained CNN. 

Low-rank methods have also been exploited at the training stage.
In \cite{denil2013predicting}, low-rank methods were exploited to reduce the number of network parameters that are learned during training. 
Low-rank methods have also been adopted to learn separable filters to accelerate
convolution in \cite{rigamonti2013learning,Amos2015Learning}, which is achieved by adding additional regularization to find low-rank filters. 

Besides low-rank matrix factorization, another way to reduce the number of parameters of weight matrix is leveraging structured matrices which can describe  $m\times n$ matrices with much fewer parameters than $mn$. In this way, Sindhwani \textit{et al.} \cite{sindhwani2015structured} proposed to learn structured parameter matrices of DNNs, which also accelerates inference and training dramatically via fast matrix-vector products and gradient computation. The work in \cite{cheng2015exploration} proposed to impose the circulant structure on the weight matrix of fully-connected layers to accelerate computation both at training and inference stages. In \cite{yang2015deep}, the authors presented an adaptive Fastfood transform to reparameterize the matrix-vector multiplication of fully-connected layers, thereby reducing the storage and computational costs.

\end{itemize}

\section{Communication-Efficient Edge AI Systems}
Due to the limited computation, storage, and communication resources of edge nodes, as well as the privacy, security, low-latency, and reliability requirements of AI applications, a variety of edge AI system architectures have been proposed and investigated for efficient training and inference. This section gives a comprehensive survey of different edge AI systems and topics therein. It starts with a general discussion on different architectures, and then introduces them one by one.

\subsection{Architectures of Edge AI Systems}
We summarize the main system architectures of edge AI into four categories. According to the availability data and model parameters, {data partition based edge training systems} and {model partition based edge training systems} are two common system architectures for efficiently training at network edges. To achieve low-latency inference, {computation offloading based edge inference systems} is a promising approach by offloading the entire or a part of inference tasks from resource limited edge devices to proximate edge servers. There are also edge AI systems defined by general computing paradigms, which can be termed as {general edge computing systems}.
\begin{itemize}[leftmargin=0pt,itemindent=1.5em,align=left,topsep=0.5em,itemsep=0.5em]\setlength{\parindent}{1em}
    \item \textbf{Data partition based edge training systems:} For data partition based edge training systems, the data is massively distributed over a number of edge devices, and each edge device has only a subset of the whole dataset. Then the edge AI model can be trained by pooling the computation capabilities of edge devices. During training, each edge device holds a replica of the complete AI model to compute a local update. This procedure often requires a centralized coordinating center, e.g., an edge server, for scheduling a number of edge devices, aggregating the local updates from edge devices, etc. There are also works considering decentralized systems where edge devices communicate with each other directly. Edge training systems with a central node are usually called \textit{distributed} system modes, while systems without a central node are called \textit{decentralized} system modes, as demonstrated in Fig. \ref{fig:system_mode}.
    \begin{figure}[h]
      \centering
      \subfigure[Distributed mode]{\includegraphics[width=0.48\columnwidth]{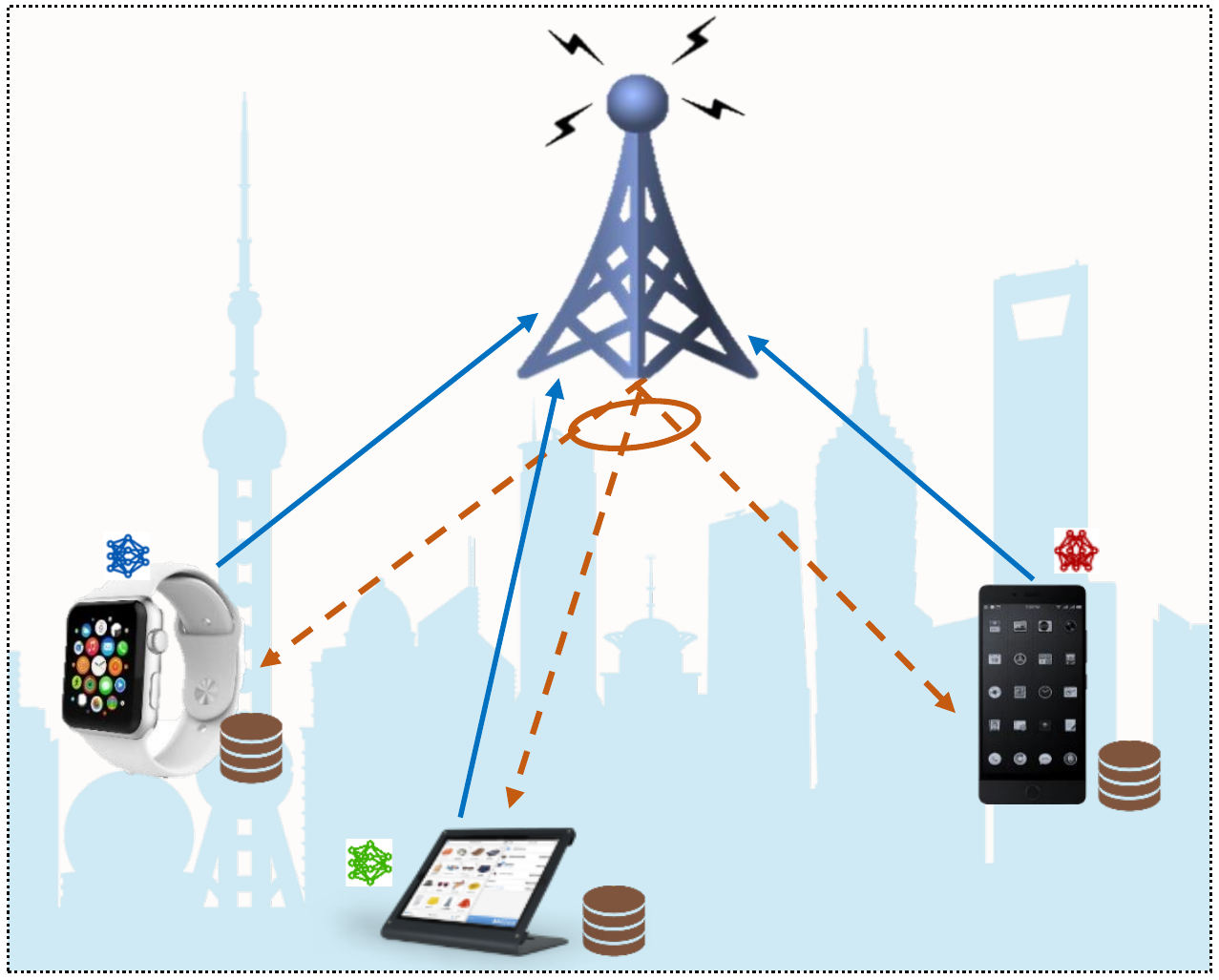}\label{fig:distributed}}
      \subfigure[Decentralized mode]{\includegraphics[width=0.48\columnwidth]{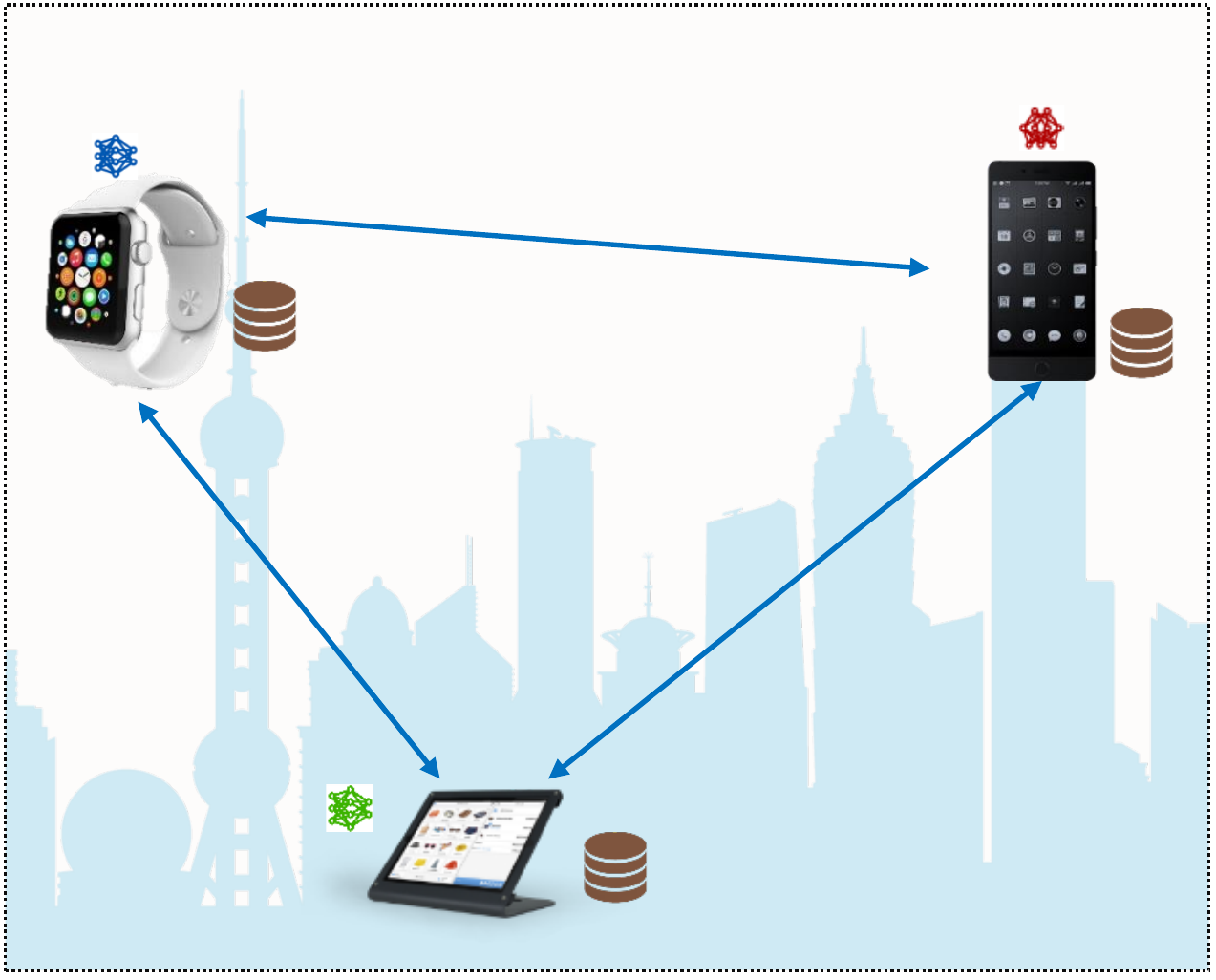}\label{fig:decentralized}}
      \caption{Two different types of edge training systems.}
      \label{fig:system_mode}
    \end{figure}

    \item \textbf{Model partition based edge training systems:} In model partition based edge training systems, each node does not have replica of all the model parameters, i.e., the AI model is partitioned and distributed across multiple nodes. Model partition is needed when very deep machine learning models are applied. Some works proposed to balance computation and communication overhead via model partition for accelerating the training process. Furthermore, model partition based edge training systems garner much attention for preserving the data privacy during training when each edge node can only access to partial data attributes for a common set of user identities. It is often referred to as \textit{vertical federated learning} \cite{yang2019federated}. To preserve data privacy, it is proposed to train a model through the synergy of the edge device and edge server by performing simple processing at the device and uploading the intermediate values to a powerful edge server. This is realized by deploying a small part of model parameters on the device and the remaining part on the edge server to avoid the exposure of users' data.

    \item \textbf{Computation offloading based edge inference systems:} To enable low-latency edge AI services, it is critical to deploy the trained model proximate to end users. Unfortunately, it is often infeasible to deploy large models, especially DNN models, directly on each device for local inference due to the limited storage, computation and battery resources. Therefore, a promising solution is to push the AI model and massive computations to proximate edge servers, which prompts the recent proposal of computation offloading based edge inference systems \cite{Khaled_MEC17}. We divide the works on computation offloading based edge inference systems into two classes, i.e., deploying the entire model on an edge server, and partitioning the model and deploying across the edge device and edge server.

    \item \textbf{General edge computing systems:} Beyond the systems mentioned above, there are also edge AI systems defined by general computing paradigms, e.g., MapReduce \cite{dean2008mapreduce}. The MapReduce-like frameworks often consider distributed data input and distributed model deployment jointly for accelerating distributed training or inference. In such systems, reducing the communication overhead for data shuffling between multiple nodes becomes a critical task. Interestingly, coding technique plays a critical role in 
    scalable data shuffling \cite{Ali_CCT16,Yuanming_WDCTSP18} as well as straggler mitigation \cite{parrinello2018coded}.
\end{itemize}
In the remaining part of this section, we discuss the important topics and involved techniques that address the communication challenges for these system architectures.

\subsection{Data Partition Based Edge Training Systems}
In data partition based edge training systems, each device usually has a subset of the training data and a replica of the machine learning model. The training can be accomplished by performing local computation and periodically exchanging local updates from mobile devices. The main advantage of such a system is that it is applicable to most of the model architectures and scales well. The main drawback is that the model size and the operations that are needed to complete the local computation are limited by the storage size and computation capabilities of each device. In the following, we separately discuss distributed and decentralized system modes.

\subsubsection{Distributed System Mode}
In the distributed system mode, each edge device computes a local update according to its local data samples, and the central node shall periodically aggregate local updates from edge devices. The communication bottleneck comes from aggregating the local updates from mobile devices and straggler devices. The efforts for addressing the communication challenges in distributed data partition based training systems are listed as follows:
\begin{itemize}[leftmargin=0pt,itemindent=1.5em,align=left,topsep=2.5em,itemsep=0.5em]\setlength{\parindent}{1em}
\item \textbf{Fast aggregation via over-the-air computation:}
\begin{figure*}[htb]
  \centering
  \includegraphics[width=0.8\textwidth]{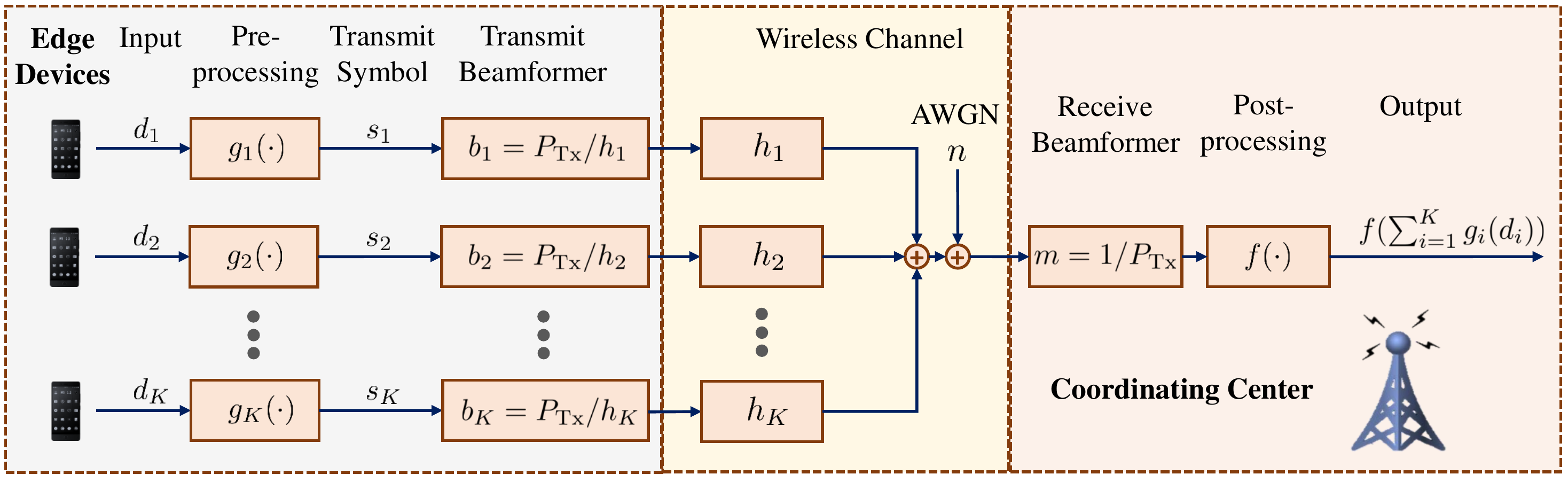}
  \caption{Illustration of fast aggregation via over-the-air computation. We exemplify it with a simple single-antenna system for computing $f(\sum_{i=1}^{K}g_i(d_i))$ from $K$ distributed edge devices.}
  \label{fig:aircomp}
\end{figure*}
Over-the-air computation is an efficient approach to compute a function of distributed data by exploiting the signal superposition property of the wireless multiple access channel \cite{nazer2007computation}. As shown in Fig. \ref{fig:aircomp}, we are able to jointly consider communication and computation to reduce the communication costs significantly. 
In particular, the function that can be computable via over-the-air computation is called the nomographic function \cite{Goldenbaum_TSP13harnessing}. In distributed machine learning, we first compute the local updates (e.g., gradients and model parameters) at each worker, and aggregate these values over the wireless channel. For aggregation functions that fall into the class of nomographic functions, we are able to improve the communication efficiency by exploiting over-the-air computing. It should be noted that  digital modulation schemes for over-the-air computation are advocated in \cite{wu2020noma,chang2020communication,dong2020distributed,zhu2020one} due to its easier implementation on the existing communication systems and its less stringent requirement of synchronization compared to anolog schemes.

To improve the communication efficiency for federated learning, Yang \textit{et al.} \cite{yang2018federated} proposed to adopt the over-the-air computation approach for fast model aggregation instead of the traditional communication-and-computation separation method.  This is motivated by the fact that the aggregating function is a linear combination of updates from distributed mobile devices, which falls into the set of nomographic functions. Using transceiver design by exploring the signal superposition property of a wireless multiple access channel, over-the-air computation can improve the communication efficiency and reduce the required bandwidth. In addition, the joint device selection and beamforming design problem was considered in \cite{yang2018federated}, for which sparse and low-rank optimization methods were proposed, yielding admirable performance of the proposed over-the-air computation for fast model aggregation. 

The efficiency of over-the-air computation for fast aggregation in federated edge learning has also been demonstrated in  \cite{zhu2018low}, which characterized two trade-offs between communication and learning performance. The first one is the trade-off between the updated quality measured by the receive SNR  and the truncation ratio of model parameters due to the proposed truncated-channel-inversion policy for deep fading channels. The second one is the trade-off between the receive SNR and the fraction of exploited data, namely, the fraction of scheduling cell-interior devices if the data distributed over devices uniformly. In \cite{Kaibin_arXiv18},  over-the-air computation based on MIMO, i.e., multi-antenna techniques, is further adopted in high-mobility multi-modal sensing for fast aggregation, where the receive beamforming is designed based on the differential geometry approach.

Based on over-the-air computation, Amiri and Gunduz \cite{amiri2019machine} proposed a gradient sparsification and random linear projection method to reduce the dimension of gradients due to limited channel bandwidth. It was shown that such an approach results in a much faster convergence of the learning process compared with the separate computation and
communication based approaches. This work was further extended to wireless fading channels in  \cite{amiri2019federated}.

\item \textbf{Aggregation frequency control with limited bandwidth and computation resources:}
The learning process includes the local updates at different devices and the global aggregation at the fusion center. We can aggregate the local updates at the interval of one  or multiple local updates, such as adopting the federated averaging algorithm \cite{mcmahan2017communication}. The aggregation frequency should be carefully designed by weighing the limited computation resources at devices locally and the limited communication bandwidth for global data aggregation. To this end, Wang \textit{et al.} \cite{Kevin_arXiv18federated} provided a convergence bound of gradient-descent based federated learning from a theoretical perspective. Based on this convergence result, the authors proposed a control algorithm that learns the data distribution, system dynamics, and model characteristics, which can be used to dynamically determine the frequency of global aggregation in real time to minimize the learning loss under a fixed resource budget. Zhou and Cong \cite{zhou2018convergence} established the convergence results of the distributed stochastic gradient descent algorithm that is averaged after \(K\) steps for nonconvex loss functions. The convergence rate in terms of the total run time instead of the number of iterations was investigated in \cite{wang2018adaptive}, which also proposed an adaptive communication strategy that starts with a low aggregation frequency to save communication costs, followed by increasing the aggregation frequency to achieve a low error floor.

\item 
\textbf{Data reshuffling via index coding and pliable index coding:}
Data reshuffling \cite{recht2013parallel,gurbuzbalaban2015random} is a recognized approach to improve the statistical performance of machine learning algorithms. Randomly reshuffling the training data at each device makes the distributed learning algorithm go over the data in a different order, which brings statistical gains for non-IID data \cite{attia2019near}. However, in edge-AI systems, its communication cost is prohibitively expensive. There are a sequence of works focusing on reducing the communication cost of data reshuffling.

To reduce the communication cost of data reshuffling, Lee \textit{et al.} \cite{Ramchandran_DML18} proposed a coded shuffling approach based on index coding. This approach assumes that the data placement rules are pre-specified. The statistical learning performance can be improved provided a small number of new data points updated at each work, which motivates the proposal of a pliable index coding based semi-random data reshuffling approach \cite{song2017pliable} for more efficient coding schemes design. It claims that the new data for each device is not necessarily in a specific way and each data is required at no more than $c$ devices (which is called the $c$-constraint). The pliable data reshuffling problem was also considered in wireless networks \cite{jiang2019pliable}. It was further observed that at per round it is not necessary to update a new data for all mobile devices, and the authors proposed to maximize the number of devices that are refreshed with a new data point. This approach turns out to reduce the communication cost considerably with a slight sacrifice of the learning performance.

\item 
\textbf{Straggler mitigation via coded computing:}
In practice, some devices may be stragglers during the computation of the gradients, i.e., it takes more time for these devices to finish the computation task.  
By carefully replicating data sets on devices, Tandon \textit{et al.} \cite{Tandon_pmlr-v70-tandon17a} proposed to encode the computed gradients to migrate stragglers, while the amount of redundancy data depends on the number of stragglers in the system. In \cite{ye2018communication},  straggler tolerance and communication cost were considered jointly. Therefore, compared with \cite{Tandon_pmlr-v70-tandon17a}, the total runtime of the distributed gradient computation is further reduced by distributing the computations over subsets of gradient vector components in addition to subsets of data sets. Raviv \textit{et al.} \cite{raviv2018gradient} adopted tools from classic coding theory, i.e.,  cyclic MDS codes, to achieve favorable performance of gradient coding in terms of the applicable range of parameters and in the complexity of the coding algorithms. Using Reed-Solomon codes, Halbawi \textit{et al.} \cite{halbawi2018improving} made the learning system more robust to stragglers compared with \cite{Tandon_pmlr-v70-tandon17a}.  The performance with respect to the communication load and computation load required for mitigating the effect of stragglers was further improved in  \cite{li2018near}.
Most of straggling mitigation approaches assumed that the straggler devices have no contribution to the learning task. In contrast, it was proposed by \cite{ozfaturay2018speeding} to exploit the non-persistent stragglers since they are able to complete a certain portion of assigned tasks in practice.
This is achieved by transmitting multiple local updates from devices to the fusion center per communication round instead of only one local updates per round.

In addition, approximate gradient coding was proposed in \cite{raviv2018gradient} where the fusion center only requires an approximate computation of the full gradients instead of an exact one, which  reduces the  computation from the devices significantly while preserving the system tolerance to stragglers.
However, this approximate gradient approach typically  results in a slower convergence rate of the learning algorithm compared with the exact gradient approach \cite{bitar2019stochastic}.
When the loss function is the squared loss, it was proposed in \cite{maity2019robust} to encode the second moment of the data matrix with  a low density parity-check (LDPC) code to mitigate the effect of the stragglers. They also indicated that the moment encoding based gradient descent algorithm can be viewed as a stochastic gradient descent method, which provides opportunities to obtain convergence guarantees for the proposed approach. Considering the general loss function, it was proposed in \cite{horii2019distributed} to distribute the data to the devices using  low density generator matrix (LDGM) codes. Bitar \textit{et al.}
\cite{bitar2019stochastic} proposed an approximate gradient coding scheme by distributing data points redundantly to devices based on a pair-wise balanced design, simply ignoring the stragglers. The convergence guarantees are established and the convergence rate can be improved with the redundancy of data \cite{bitar2019stochastic}.
\end{itemize}

\subsubsection{Decentralized System Mode}
In the decentralized mode, a machine learning model is trained with a number of edge devices by exchanging information directly without a central node. A well known decentralized information exchange paradigm is the gossip communication protocol \cite{boyd2006randomized}, by randomly evoking a node as a central node to collect updates from neighbour nodes or broadcast its local update to neighbour nodes. By integrating the gossip communication protocols into the learning algorithms, Elastic Gossip \cite{pramod2018elastic} and Gossiping SGD \cite{daily2018gossipgrad} \cite{blot2019distributed} \cite{blot2016gossip} were proposed. 

One typical network topology for decentralized machine learning is the fully connected network, where each device communicates directly with all other devices. In this scenario, each device maintains a local copy of the model parameters and computes its local gradients that will be sent to all other devices. Each device can average the gradients received from every other devices and then perform local updates. In each iteration, the model parameters will be identical at all devices if each device starts from a same initial point.  This process is essentially the same as the classical gradient descent at a centralized server, so the convergence can be guaranteed as in the centralized settings. However, such a fully connected network suffers a heavy communication overhead that grows quadratically in the number of devices, while the communication overhead is linear in the number of devices for centralized settings. Therefore, network topology design plays a key role in alleviating the communication bottleneck in decentralized scenarios. In addition, the convergence rate of the decentralized algorithm also depends on the topology of network \cite{nedic2018network}. We should note that the decentralized edge AI system suffers from the same issues as the system in distributed mode since each device acts like a fusion center. 

There have been several works demonstrating that some carefully designed topologies of networks achieve better performance than the fully connected network.
It has been empirically observed in \cite{adjodah2019communication} that using an alternative network topology between devices can lead to improved learning performance in several deep reinforcement learning tasks compared with the standard fully-connected communication topology. Specifically, it was observed in \cite{adjodah2019communication} that the Erdos-Renyi graph topology with 1000 devices can compete with
the standard fully-connected topology with 3000 devices, which shows that the machine learning performance  can be more efficient if the topology is carefully designed. Considering that different devices may require different times to carry out local computation, Neglia \textit{et al.} \cite{neglia2019role} analyzed the influences of different network topologies on the total runtime of distributed subgradient methods, which can determine the degrees of the topology graph, leading to the faster convergence speed. They also showed that a sparser network can sometimes result in significant reduction of the convergence time.

One common alternative to the fully connected network topology is to employ a ring topology \cite{patarasuk2009bandwidth}, where each device only communicates with its neighbors that are arranged in a logical ring. 
More concretely, each device aggregates and passes its local gradients along the ring such that all devices have a copy of the full gradients at the end. This approach has been adopted in distributed deep learning for model updating \cite{jin2016scale,sergeev2018horovod}. However, the algorithm deployed on the ring topology are inherently sensitive to stragglers \cite{reisizadeh2019codedreduce}.
To alleviate the effects of stragglers in the ring topology, Reisizadeh \textit{et al.} 
\cite{reisizadeh2019codedreduce} proposed to use a logical tree topology for communication, based on which they mitigated stragglers by gradient coding techniques. In the tree topology, there are several layers of devices, where each device communicates only with its parent node. By concurrently transmitting messages from a large number of children nodes to multiple parent nodes, communication with the tree topology can be more efficient than that with the ring topology.

\subsection{Model Partition Based Edge Training Systems} 
While data partition based edge training systems have obtained much attention in both academia and industry, there is also an important line of works designing edge AI systems based on partitioning a single machine learning model and deploying it distributedly across mobile devices and edge servers. In such systems, each node holds part of the model parameters and accomplish the model training task or the inference task collaboratively.
One main advantage of model partition in the training process is the small storage size needed for each node. In this system, the machine learning model is distributedly deployed among multiple computing nodes, with each node evaluating updates of only a portion of the model's parameters. Such method is particularly useful in the scenarios where the machine learning model is too large to be stored in a single node \cite{mayer2019scalable,xing2015petuum}. Another main concern of model partition during training is the data privacy when the data at each node belongs to different parties. However, model training with model partition based architectures also poses heavy communication overhead between edge devices. 

\begin{itemize}[leftmargin=0pt,itemindent=1.5em,align=left,topsep=0.5em,itemsep=0.5em]\setlength{\parindent}{1em}
\item \textbf{Model partition across a large number of nodes to balance computation and communication:}
A line of works \cite{Dean_ICML17placement,harlap2018pipedream,huang2019gpipe} have considered the model partition across edge nodes with heterogeneous hardware and computing power. In \cite{Dean_ICML17placement}, a reinforcement learning approach was proposed for deploying the computing graph onto edge computing devices, which, however, is time and resource intensive.
To avoid the huge computation cost of reinforcement learning based approach, Harlap \textit{et al.} \cite{harlap2018pipedream} proposed the PipeDream system for automatically determining the model partition strategy of DNNs. Furthermore, injecting multiple mini-batches makes the system converge faster than using a single machine or using the data partition approach.  
While PipeDream stresses the hardware utilization of edge devices, each device should maintain multiple versions of model parameters to avoid optimization issues caused by the staleness of parameters with asynchronous backward updates. This hinders scaling to much bigger models for PipeDream. To address this problem, the GPipe system was proposed in \cite{huang2019gpipe} with novel batch-splitting and re-materialization techniques, which is able to scale to large models with little additional communication overhead. 

\item  \textbf{Model partition across the edge device and edge server to avoid the exposure of users' data:}
In practice, powerful edge servers are often owned by service providers, but users may be relunctant to expose their data to service providers for model training. The observation that a DNN model can be split between two successive layers motivates researchers to deploy the first few layers on the device locally and the remaining layers on the edge server to avoid the exposure of users' data. Mao \textit{et al.} \cite{mao2018privacy} proposed a privacy-preserving deep learning architecture where the shallow part of a DNN is deployed on the mobile device and the large part is deployed on the edge server. Gupta and Raskar \cite{gupta2018distributed} designed a model partition approach over multiple agents, i.e., multiple data sources and one supercomputing resource, and further extended it to semi-supervised learning cases with few labeled sample. A particular DNN model for face recognition is trained and evaluated on a Huawei Nexux 6P phone with satisfactory performance. In \cite{wang2018not}, a partition approach named ARDEN was proposed by taking both privacy and performance into consideration. The model parameters at mobile device are fixed and differential privacy mechanism is introduced to guarantee the privacy of the output at mobile device. Before uploading the local output, deliberate noise is added to improve the robustness of DNN, which is shown to be beneficial for the inference performance.

\item \textbf{Vertical architecture for privacy with vertically partitioned data and model:}
In most industries, data is often vertically partitioned, i.e., each owner only holds partial data attributes.   
Data isolation becomes a severe bottleneck for collaboratively building a model due to competition, privacy, and administrative procedures. Therefore, much attention is being paid on privacy-preserving machine learning with vertically partitioned data \cite{yang2019federated}. During the training, the model is also vertically partitioned and each owner holds a part of model parameters. Therefore, \textit{vertical architecture} of AI is proposed and studied for  privacy-preserving machine learning where each node has access to different features of common data instances and maintains the corresponding subset of model parameters. What makes it worse is that the label of each data instance is only available to nodes belonging to one party.

Vaidya and Clifton \cite{vaidya2003privacy} proposed a privacy-preserving $k$-means algorithm in the vertical architecture with secure multi-party computation. Kantarcioglu and Clifton \cite{kantarcioglu2004privacy} studied the secure association rules mining problem with vertically partitioned data. A linear regression model was taken into consideration in \cite{gascon2016secure}, and multi-party computation protocols were proposed with a semi-trusted third party to achieve secure and scalable training.
For privacy-preserving classification with support vector machine (SVM), Yu \textit{et al.} \cite{yu2006privacy_svm} considered the dual problem of SVM and adopted a random perturbation strategy, which is suitable only for nodes belong to more than three parties. A privacy-preserving classification approach based on decision tree was proposed in \cite{vaidya2005privacy}, which adopts secure multi-party computation procedures including commutative encryption to determine if there are any remaining attributes and secure cardinality computation of set intersection. For classification with logistic regression, the problem becomes even more difficult because of the coupled objective function as well as the gradient. To address this problem, Hardy \textit{et al.} \cite{hardy2017private} proposed to use Taylor approximation to benefit from the homomorphic encryption protocol without revealing the data at each node.
\end{itemize}


\subsection{Computation Offloading Based Edge Inference Systems}
\begin{figure}[h]
      \centering
      \subfigure[Server-based edge inference]{\includegraphics[width=\columnwidth]{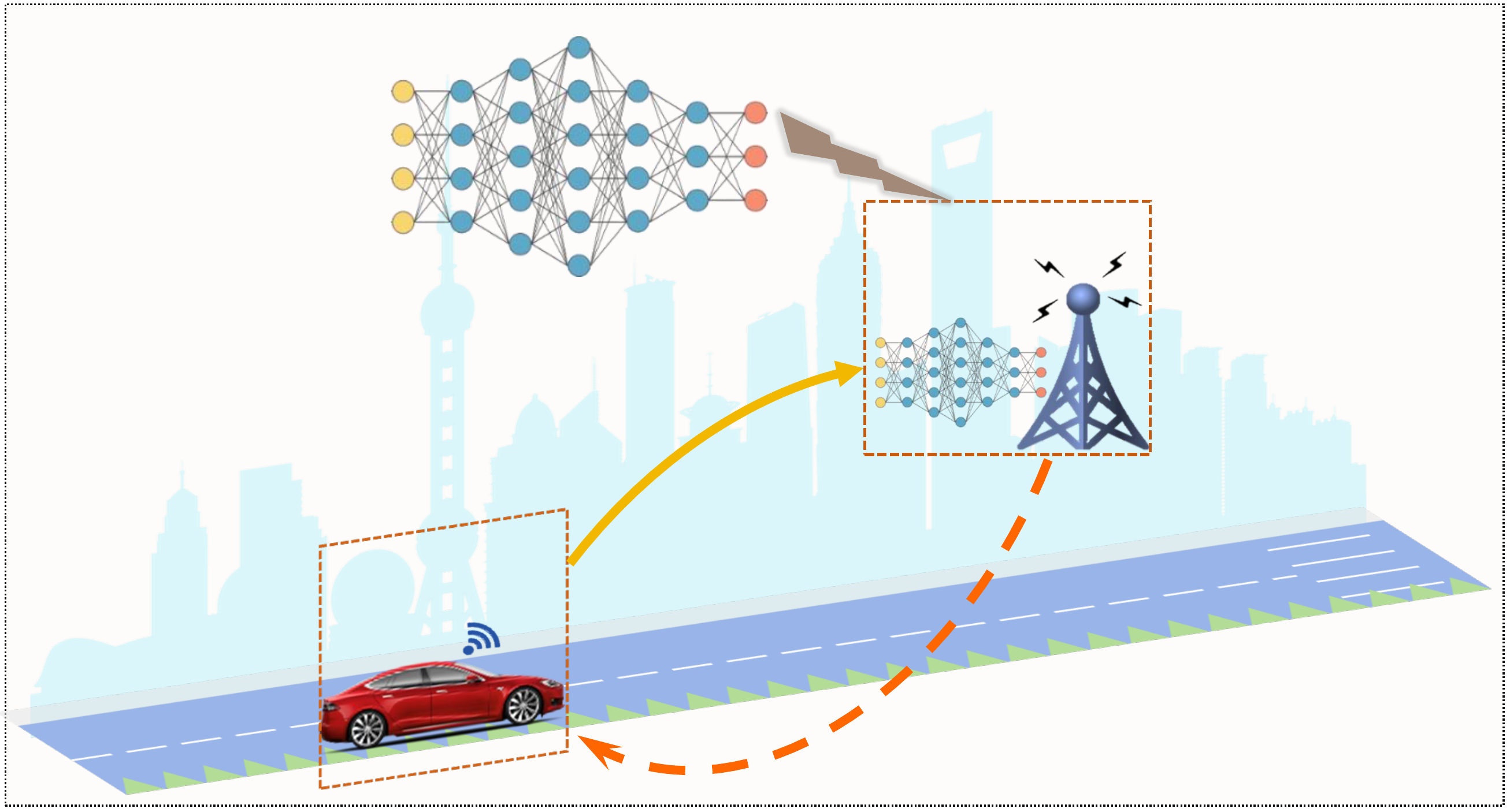}\label{fig:offloading_entire}}
      \subfigure[Device-edge joint inference]{\includegraphics[width=\columnwidth]{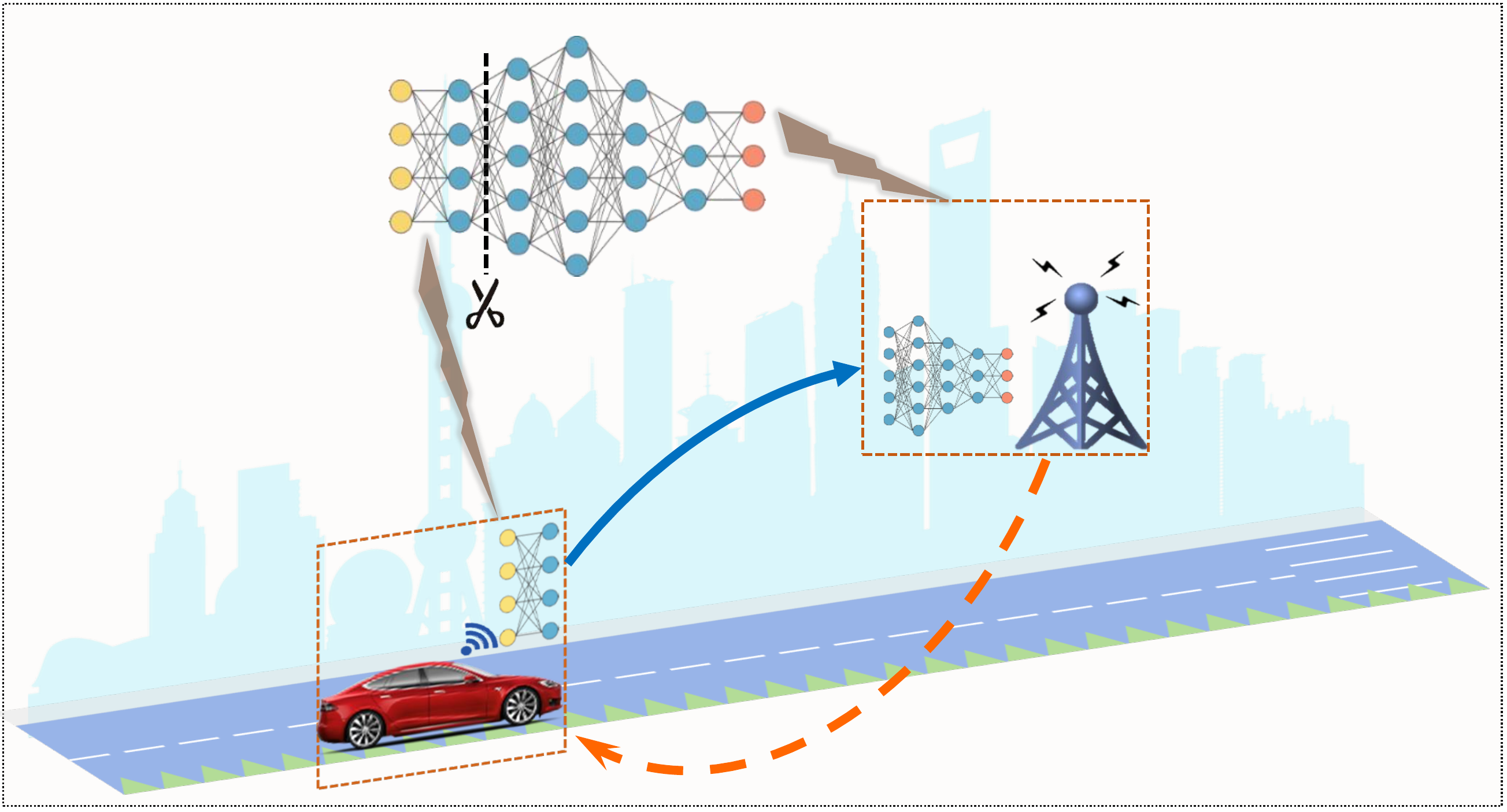}\label{fig:offloading_partition}}
      \caption{Computation offloading based edge inference systems.}
      \label{fig:offloading}
    \end{figure}

The advancement of edge computing makes it increasingly attractive to push the AI inference task to network edge to enable low-latency AI services for mobile users \cite{Khaled_MEC17}. However, the power consumption and storage for DNN models is often unbearable for mobile devices such as wearable devices. Fortunately, offloading the task from edge devices to powerful edge servers emerges as an antidote \cite{Khaled_MEC17,yang2019energy,hua2019reconfigurable,teerapittayanon2017distributed}. One solution is to offload the entire inference task to an edge server, which is termed as \textit{server-based edge inference}, as shown in Fig. \ref{fig:offloading_entire}. It is particularly suitable for resource limited IoT devices. In this case, the entire AI models are deployed on edge servers and edge devices should upload their input data to edge servers for inference. For latency and privacy concerns, another alternative is offloading only a part of the task to the edge server, and the edge server computes the inference result based on the intermediate value computed by the edge device. We refer to it as \textit{device-edge joint inference} as shown in Fig. \ref{fig:offloading_partition}. This edge device and edge server synergy can be achieved by performing simple processing at the device and the remaining part at the edge server. 

\subsubsection{Server-Based Edge Inference}
In the scenario where the models are deployed on the edge servers, the devices send the input data to the edge, the edge servers compute the inference results according to the trained models, and the inference results are then sent back to the devices. The main bottleneck is the limited communication bandwidth for data transmission. 
To reduce the real-time data transmission overhead of uplink transmission in bandwidth-limited edge AI systems, an effective way is to reduce the volume of data transmitted from devices without hurting the inference accuracy. In addition, cooperative downlink transmission of multiple edge servers has been proposed to enhance the communication efficiency for edge inference.
\begin{itemize}[leftmargin=0pt,itemindent=1.5em,align=left,topsep=0.5em,itemsep=0.5em]\setlength{\parindent}{1em}
    \item 
    \textbf{Partial data transmission:}
    To realize cloud based visual localization for mobile robots in real time, it is important to control the volume of the data through the network. Therefore, Ding \textit{et al.} \cite{ding2019communication} used an data volume reduction method proposed by  \cite{paull2015communication} for multi-robot communication, which employs sparsification methods to compress the data.
    In a cloud-based collaborative 3D mapping system, Mohanarajah \textit{et al.} \cite{mohanarajah2015cloud} proposed to reduce bandwidth requirements by sending only the key-frames as opposed to  all the frames produced by the sensor, and Chen \textit{et al.} \cite{chen2015glimpse} proposed to determine and offload key-frames for object detection by utilizing heuristics such as frame differences to select the key-frames. These approaches are useful in reducing the communication cost when we are able to exploit the structure of the specific tasks and the associated data.

    \item 
    \textbf{Raw data encoding:}
    Data encoding has been widely used in compressing the data volume. For example, traditional image compression approaches (e.g., JPEG) can compress data aggressively, but they are often optimized from the perspective of human-visual, which will result in an unacceptable performance degradation in DNN applications if we use a high compression ratio. Based on this observation, to achieve a higher compression ratio, Liu \textit{et al.} \cite{liu2018deepn} proposed to optimize the data encoding schemes from the perspective of DNNs based on the frequency component analysis and rectified quantization table, which is able to achieve a higher compression ratio than the traditional JPEG method without degrading the accuracy for image recognition.  Instead of using standard video encoding techniques, it was argued in \cite{chinchali2018neural} that data collection and transmission schemes should be designed jointly in vision tasks to maximize an end-to-end goal with a pre-trained model. Specifically, the authors proposed to use DNNs to encode the high dimensional raw data into a sparse, latent representation for  efficient transmission, which can be recovered later at the cloud via a decoding DNN.  In addition, this coding process is controlled by a reinforcement learning algorithm, which sends action information to devices for encoding in order to maximize the predication accuracy  of the pre-trained model with decoded inputs, while achieving communication-efficient data transmission.
    This novel data encoding idea is a promising solution for realizing real-time inference in edge AI systems.
    \item \textbf{Cooperative downlink transmission:} Cooperative transmission \cite{gesbert2010multi} is known as an effective aproach for improving the communication efficiency via proactive interference-aware coordination of multiple base stations. It was proposed in \cite{yang2019energy} to offload each inference task to multiple edge servers and cooperatively transmit the output results to mobile users in downlink transmission. A new technology named intelligent reflecting surface (IRS) \cite{qingqing2019towards} emerges as a cost-effective approach to enhance the spectrum efficiency and energy efficiency of wireless communication networks, which is promising in facilitating communication-efficient edge inference \cite{yuan2020reconfigurableintelligentsurface}. It is achieved by reconfiguring the wireless propagation environment via a planar array to induce the change of the signals' amplitude and/or phase. To further improve the performance of the cooperative edge inference scheme in \cite{yang2019energy}, Hua \textit{et al.} \cite{hua2019reconfigurable} proposed the IRS-aided edge inference system and designed a task selection strategy to minimize both the uplink and downlink transmit power consumption, as well as the computation power consumption at edge servers.

\end{itemize}

\subsubsection{Device-Edge Joint Inference}
For many on-device data, such as healthcare information and users' behaviors, privacy is of a primary concern. Thus, there emerges the idea of edge device and edge server synergy, which can be termed as \textit{device-edge joint inference}, by deploying the partitioned DNN model over the mobile device and the powerful edge server. By deploying the first few layers locally, a mobile device can compute the local output with simple processing, and transmit the local output to a more powerful edge server without revealing any sensitive information.

An early work \cite{hauswald2014hybrid} considered the partition of the image classification pipeline and found that executing feature extraction on devices and offloading the rest to the edge servers achieves optimal runtime. Recently, Neurosurgeon has been proposed in \cite{kang2017neurosurgeon}, where a DNN model is automatically split between a device and an edge server according to the network latency for transmission and the mobile device energy consumption at different partition points, in order to minimize the total inference time. Different methods have been developed \cite{mao2017modnn,zhao2018deepthings} to  partition a pre-trained DNN over several mobile devices in order to accelerate DNN inference on devices. Bhardwaj \textit{et al.} \cite{bhardwaj2019memory} further considered memory and communication costs in this distributed inference architecture, for which model compression and network science-based knowledge partitioning algorithm are proposed to address these issues. For robotics system where the model is partitioned between the edge server and the robot, the robot should take both local computation accuracy and offloading latency into account, and this offloading problem was formulated in \cite{chinchali2019network} as a sequential decision making problem that is solved by a deep reinforcement learning algorithm.

In the following, we review the main methods for further reducing the communication overhead for the model partition based edge inference.
\begin{itemize}[leftmargin=0pt,itemindent=1.5em,align=left,topsep=0.5em,itemsep=0.5em]\setlength{\parindent}{1em}
\item 
\textbf{Early exit:}
Early exit can be used to reduce communication workloads when partitioning DNNs, which has been proposed in \cite{teerapittayanon2016branchynet} based on the observation that the features learned at the early layer of the network can be often sufficient to produce accurate  inference results. Therefore, the inference process can exit early if the data samples can be inferred with high confidence. This technique has been adopted in \cite{teerapittayanon2017distributed} for distributed DNN inference over the cloud, the edge and devices. With early exit, each device first performs the first few layers of an DNN, and offloads the rest of computation to the edges or the clouds if the outputs of the device do not meet the accuracy requirements. This approach is able to reduce the communication cost by a factor of over \(20\times\) compared with the traditional approach that offloads all raw data to the cloud for inference. More recently, Li \textit{et al.} \cite{li2019edge} proposed an on-demand low-latency inference framework through jointly designing the model partition strategy according to the heterogeneous computation capabilities between a mobile device and edge servers, and the early exit strategy according to the complicated network environment.

\item \textbf{Encoded transmission and pruning for compressing the transmitted data:} 
In a hierarchical distributed architecture, the main communication bottleneck is that the transmission of intermediate values between the partition point since the intermediate data can be much larger than the raw data.
To reduce the communication overhead of intermediate value transmissions, it was proposed in 
\cite{ko2018edge} to partition a network at an intermediate layer, whose features are encoded before wireless transmissions to reduce their data size. It shows that partitioning a
CNN at the end of the last convolutional layer where the data communication requirement is less coupled with feature space encoding enables significant reduction in communication workloads. Recently, a deep learning based end-to-end architecture was proposed in \cite{shao2019bottlenet++}, named BottleNet++. By jointly considering model partition, feature compression and transmission, BottleNet++ achieves up to 64x bandwidth reduction over the additive white Gaussian noise channel and up
to 256x bit compression ratio in the binary erasure channel, with less than $2\%$ reduction in accuracy, compared with merely transmitting intermediate data without feature compression.

Network pruning, as discussed in Section \ref{subsec:pruning}, has been exploited in reducing the communication overhead of intermediate feature transmissions. For example, a 2-step pruning approach was proposed in \cite{shi2019improving} to reduce the transmission workload at the network partition point by limiting the pruning region. Specifically, the first step aims to  reduce the total computation workload of the network while the second step aims to compress the intermediate data for transmission.

\item 
\textbf{Coded computing for cooperative edge inference:} 
Coding theory can be leveraged to address the communication challenges of distributed inference in edge AI systems. For example, Zhang and Simeone \cite{zhang2019model}  considered distributed linear inference in mobile edge AI system, where the model is partitioned among several edge devices that compute the inference results cooperatively for each device. It was shown in \cite{zhang2019model} that coding is efficient in reducing  the overall computation-plus-communication latency. 
\end{itemize}

\subsection{General Edge Computing System}
Beyond the edge AI system architectures mentioned above, there are also edge AI systems based on a general computing paradigm, namely, MapReduce. MapReduce \cite{dean2008mapreduce} is a general distributed computing framework that is able to achieve parallel speedups on a variety of machine learning problems during training and inference procedures \cite{chu2007map}. The MapReduce-like distributed computing framework takes the distributed data input and distributed model deployment into account jointly. In \cite{basit2016mapreduce}, the convolutional neural network was implemented based on the MapReduce framework to accelerate its training process. Ghoting \textit{et al.} \cite{ghoting2011systemml} proposed SystemML based on the MapReduce framework to support distributed training for a broad class of supervised and unsupervised machine learning algorithms. \cite{Yuanming_WDCTSP18} proposed a communication-efficient wireless data shuffling strategy for supporting MapReduce-based distributed inference tasks.

\begin{figure}[htb]
  \centering{\includegraphics[width=\linewidth]{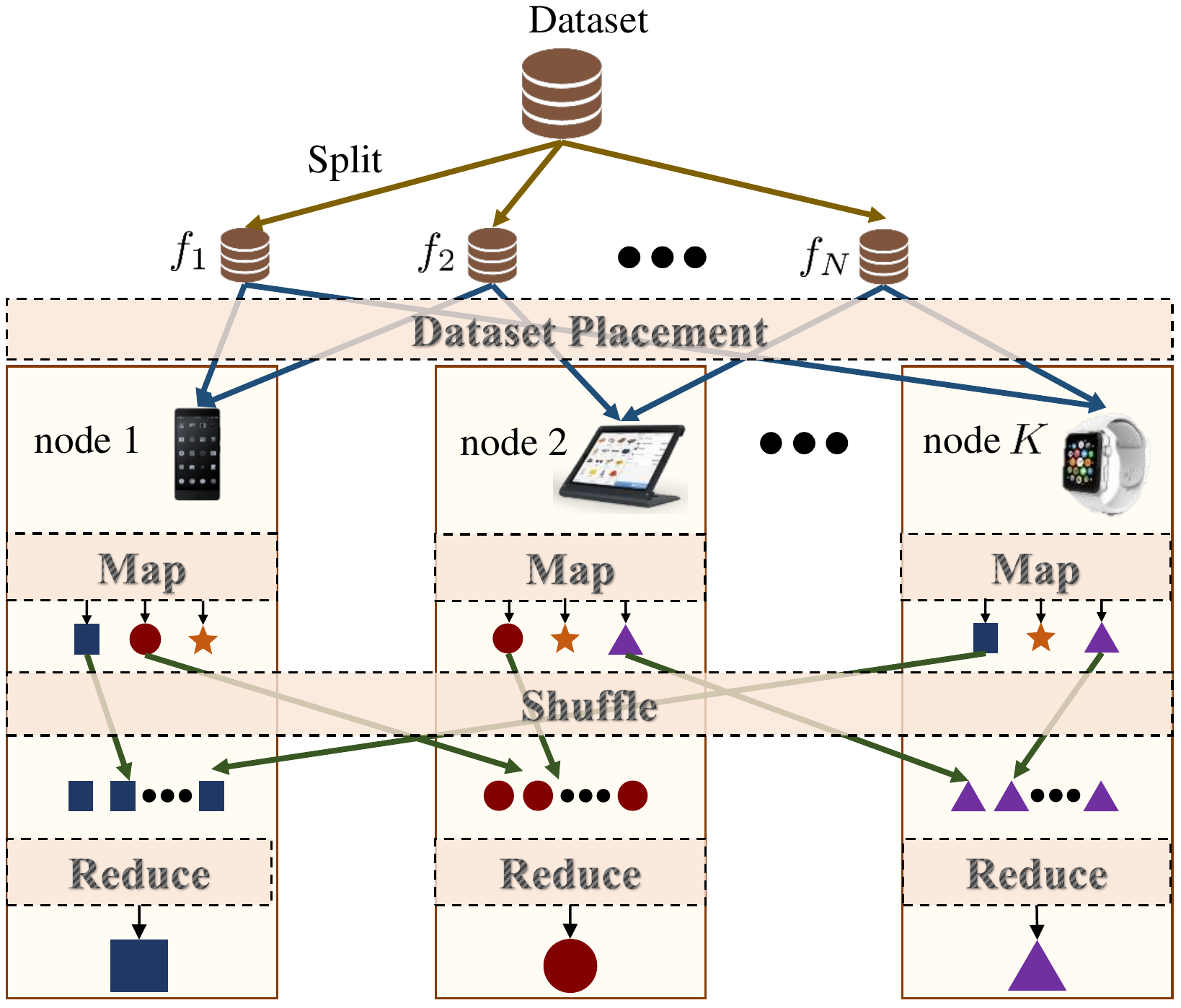}
  \caption{MapReduce computation model \cite{Yuanming_WDCTSP18}.}
  \label{fig:mapreduce}}
\end{figure} 

In the MapReduce-like distributed computing framework as shown in Fig. \ref{fig:mapreduce}, there are generally three phases (i.e., a map phase, a shuffle phase, and a reduce phase) to complete a computational task.  In the map phase, every computing node computes a map function of the assigned data simultaneously, generating a number of intermediate values. In the shuffle phase, nodes communicate with each other to obtain some intermediate values for computing the output function. Subsequently, in the reduce phase, each node computes the assigned output function according to the available intermediate values. However, there are two main bottlenecks in such a distributed computing framework. One is the heavy communication load in the shuffle phase, and another is the straggler delay caused by the variability of computation times at different nodes. To address these problems, coding has been proposed as a promising approach by exploiting abundant computing resources at the network edge \cite{Ali_ComMag17}. In recent years, coding techniques are becoming a hot area of research for reducing the communication cost of data shuffling, as well as reducing the computing latency by mitigating straggler nodes, as reviewed below.


\begin{itemize}[leftmargin=0pt,itemindent=1.5em,align=left,topsep=0.5em,itemsep=0.5em]\setlength{\parindent}{1em}
    \item \textbf{Coding techniques for efficient data shuffling:}
Coding techniques for shuffling data in the MapReduce-like distributed computing framework were first proposed in \cite{Ali_CCT16}, which considered a wireline scenario where each computing node can obtain the intermediate values from other nodes through a shared link. In \cite{Ali_arXiv16WDC}, the authors extended the work in \cite{Ali_CCT16} to a wireless setting, where the computing nodes are able to communicate with each other via an access point. A scalable data shuffling scheme was proposed by utilizing a particular repetitive pattern of placing intermediate values among devices, reducing the communication bandwidth by factors that grow linearly with the number of devices. To improve the wireless communication efficiency (i.e., achieved data rates) in the data shuffle phase,  a low-rank optimization model was proposed in \cite{Yuanming_WDCTSP18} by establishing the interference alignment condition. The low-rank model is further solved by an efficient difference-of-convex-functions (DC) algorithm. Both \cite{Ali_arXiv16WDC} and \cite{Yuanming_WDCTSP18} considered the communication load minimization problem under the wireless communication setting with a central node. 

There are also some works considering the problem of reducing the communication load in data shuffling under the wireless communication scenario without a coordinating center. That is, the computing nodes can communicate with each other through an shared wireless interference channel. For example, assuming perfect channel state information, a beamforming strategy was proposed in \cite{li2018wireless} based on side information cancellation and zero-forcing to trade the abundant computing nodes for reducing communication load, which outperforms the coded TDMA broadcast scheme based on \cite{Ali_CCT16}. This work was further extended in \cite{ha2018wireless} to consider imperfect channel state information. The paper \cite{ji2018fundamental} proposed a dataset cache strategy and a coded transmission strategy for the corresponding computing results. The goal is to minimize the communication load characterized by latency (in seconds) instead of channel uses (in bits), which is more practical in wireless networks.
In \cite{parrinello2018coded}, the authors noted that to trade abundant computation for the communication load, the computational tasks must be divided into an extremely large number of subtasks, which is impractical. Therefore, they proposed to ameliorate this limitation by node cooperation and designed an efficient scheme for task assignment. Prakash \textit{et al.} \cite{prakash2018coded} investigated coded computing for distributed graph processing systems, which improves performance significantly compared with the general MapReduce framework by leveraging the structure of graphs.

\item \textbf{Coding techniques for straggler mitigation:}
Another line of work focuses on addressing the straggler problem in distributed computing by coding techniques. Mitigating the effect of stragglers utilizing coding theory was first proposed in \cite{parrinello2018coded} for a wired network. The main idea is to leverage redundant computing nodes to perform computational subtasks, while the computation result can be correctly recovered as long as the local computation results from any desired subset of computing nodes are collected. This work was extended to wireless networks \cite{reisizadeh2017latency}, where only one local computing node can send its computation results to the fusion center at a time. The paper \cite{zhao2019node} proposed a  sub-task assignment method to minimize the total latency which is composed of the latency caused by wireless communication between different computing nodes and the fusion center and the latency caused by the variability of computation time of different devices. Most of the above works focused on linear computations (e.g., matrix multiplication). However, to realize the distributed inference on state-of-the-art machine learning algorithms (e.g., DNN), non-linear computation should be taken into consideration. As a result, the work \cite{kosaian2018learning} proposed a learning-based approach to design codes that can handle the stragglers issue in distributed non-linear computation problems.
\end{itemize}

\section{Conclusions and Future Directions}
This paper presented a comprehensive survey on the communication challenges and solutions in edge AI systems, which shall support a plethora of AI-enabled applications at the network edge. Specifically, we first summarized communication efficient algorithms for distributed training AI models on edge nodes, including zeroth-order, first-order, second-order, and federated optimization algorithms. We then categorized different system architectures of edge AI systems, including data partition based, and model partition based edge training systems. Next, we revisited works bridging the gap between computation offloading and edge inference. Beyond these system architectures, we also introduced general edge computing defined AI systems. The communication issues and solutions in such architectures were extensively discussed.

The activities and applications of edge AI are growing rapidly, and a number of challenges and future directions are listed below. 
\begin{itemize}
\item \textbf{Edge AI hardware design:} Hardware of edge nodes determines the physical limits of AI systems, for which there are a growing amount of efforts on edge AI hardware design. For example, Google edge tensor processing unit (TPU)
is designed for high-speed inference at the edge. Nvidia has rolled out Jetson TX2
for power-efficient embedded AI computing.
Nevertheless, these hardwares mainly focus on performing the entire task, especially edge inference locally. In the future, a variety of edge AI hardwares will be customized for different AI system architectures and applications.

\item \textbf{Edge AI software platforms:} The past decade has witnessed the blossom of AI software platforms from top companies for supporting cloud-based AI services. Cloud-based AI service providers are trying to include edge nodes into their platforms, though edge nodes only serve as simple extensions of cloud computing nodes currently. Google Cloud IoT,
Microsoft Azure IoT,
NVIDIA EGX,
and Amazon Web Services (AWS) IoT
are able to connect IoT devices to cloud platforms, thereby managing edge devices and processing the data from all kinds of IoT devices. 

\item \textbf{Edge AI as a service:} 
For different fields and applications, there are a variety of additional design targets and constraints, thereby requiring domain-specific edge AI frameworks. Edge AI will be a service infrastructure that integrates the computation, communication, storage and power resources at network edges to enable data-driven intelligent applications. A notable example in credit industry is FATE \cite{webankfate},
an industrial grade federated learning framework proposed by Webank. A number of federated learning algorithms \cite{cheng2019secureboost,yang2019quasi} were designed to break data isolation among institutions and to preserve the data privacy during edge training. Another representative attempt of edge AI for smart healthcare is NVIDIA Clara \cite{NVIDIAclara}, which delivers AI to healthcare and life sciences with NVIDIA's EGX edge computing platform. Since Clara features federated learning, it supports an innovative approach to collaboratively build a healthcare AI model from hospitals and medical institutions, while protecting patient data.
\end{itemize}

\section*{Acknowledgement}
We sincerely thank Prof. Zhi Ding from the University of California at Davis for insightful and constructive comments to improve the presentation of this work.

\bibliographystyle{ieeetr}
\bibliography{edgeAI}
\end{document}